\documentclass{emulateapj}
\usepackage{apjfonts}

\usepackage{lscape,amsmath,graphicx,natbib,longtable,rotating}

\newcommand{\etal}{et al.}
\newcommand{\hbeta}{H{$\beta$}}
\newcommand{\halpha}{H{$\alpha$}}

\newcommand{\CIV}{C{\sevenrm IV}}
\newcommand{\CIVwave}{C{\sevenrm IV}\,$\lambda$1549}
\newcommand{\HeII}{He{\sevenrm II}}

\newcommand{\CIII}{C{\sevenrm III]}}
\newcommand{\CIIIwave}{C{\sevenrm III]}\,$\lambda$1908}
\newcommand{\AlIII}{Al{\sevenrm III}}
\newcommand{\AlIIIwave}{Al{\sevenrm III}\,$\lambda$1857}
\newcommand{\SiIII}{Si{\sevenrm III]}}
\newcommand{\SiIIIwave}{Si{\sevenrm III]}\,$\lambda$1892}
\def\MgII{Mg\,{\sc ii}}
\def\MgIIwave{Mg\,{\sc ii}\,$\lambda$2798}

\def \OIII {[O\,{\sc iii}]}
\newcommand{\OIIIa}{[O{\sevenrm\,III}]\,$\lambda$4959}
\newcommand{\OIIIb}{[O{\sevenrm\,III}]\,$\lambda$5007}
\newcommand{\OIIIab}{[O{\sevenrm\,III}]\,$\lambda\lambda$4959,5007}
\newcommand{\NII}{[N{\sevenrm\,II}]}
\newcommand{\NIIa}{[N{\sevenrm\,II}]\,$\lambda$6548}
\newcommand{\NIIb}{[N{\sevenrm\,II}]\,$\lambda$6584}

\newcommand{\SIIa}{[S{\sevenrm\,II}]\,$\lambda$6717}
\newcommand{\SIIb}{[S{\sevenrm\,II}]\,$\lambda$6731}

   \font\sevenrm=cmr7 scaled 1000

\begin{document}

\title{Comparing Single-Epoch Virial Black Hole Mass Estimators for Luminous Quasars}

\shorttitle{VIRIAL BH MASS ESTIMATORS}


\shortauthors{SHEN \& LIU}
\author{Yue Shen\altaffilmark{1} and Xin Liu\altaffilmark{1,2}}
\altaffiltext{1}{Harvard-Smithsonian Center for Astrophysics, 60 Garden
Street, MS-51, Cambridge, MA 02138, USA} \altaffiltext{2}{Einstein Fellow}

\begin{abstract}
Single-epoch virial black hole (BH) mass estimators utilizing broad emission
lines have been routinely applied to high-redshift quasars to estimate their
BH masses. Depending on the redshift, different line estimators (\halpha,
\hbeta, \MgIIwave, \CIVwave) are often used with optical/near-infrared
spectroscopy. Here we use a homogeneous sample of $60$ intermediate-redshift
($z\sim 1.5-2.2$) SDSS quasars with optical and near-infrared spectra
covering \CIV\ through \halpha\ to investigate the consistency between
different single-epoch virial BH mass estimators. We critically compare
restframe UV line estimators (\CIVwave, \CIIIwave\, and \MgIIwave) with
optical estimators (\hbeta\ and \halpha) in terms of correlations between
line widths and between continuum/line luminosities, for the high-luminosity
regime ($L_{5100}>10^{45.4}\,{\rm erg\,s^{-1}}$) probed by our sample. The
continuum luminosities of $L_{1350}$ and $L_{3000}$, and the broad line
luminosities are well correlated with $L_{5100}$, reflecting the homogeneity
of quasar spectra in the restframe UV-optical, among which $L_{1350}$ and the
line luminosities for \CIV\ and \CIII\ have the largest scatter in the
correlation with $L_{5100}$. We found that the \MgII\ FWHM correlates well
with the FWHMs of the Balmer lines, and that the \MgII\ line estimator can be
calibrated to yield consistent virial mass estimates with those based on the
\hbeta/\halpha\ estimators, thus extending earlier results on less luminous
objects. The \CIV\ FWHM is poorly correlated with the Balmer line FWHMs, and
the scatter between the \CIV\ and \hbeta\ FWHMs consists of an irreducible
part ($\sim 0.12$ dex), and a part that correlates with the blueshift of the
\CIV\ centroid relative to that of \hbeta, similar to earlier studies
comparing \CIV\ with \MgII. The \CIII\ FWHM is found to correlate with the
\CIV\ FWHM, and hence is also poorly correlated with the \hbeta\ FWHM. While
the \CIV\ and \CIII\ lines can be calibrated to yield consistent virial mass
estimates as \hbeta\ on average, the scatter is substantially larger than
\MgII, and the usage of \CIV/\CIII\ FWHM in the mass estimators does not
improve the agreement with the \hbeta\ estimator. We discuss controversial
claims in the literature on the correlation between \CIV\ and \hbeta\ FWHMs,
and suggest that the reported correlation is either the result based on small
samples or only valid for low-luminosity objects.
\\

Based in part on observations obtained with the 6.5 m Magellan-Baade
telescope located at Las Campanas Observatory, Chile, and with the Apache
Point Observatory 3.5 m telescope, which is owned and operated by the
Astrophysical Research Consortium.
\end{abstract}
\keywords{black hole physics --- galaxies: active --- quasars: general}

\section{Introduction}\label{sec:intro}

Knowing the mass of active supermassive black holes (SMBHs) is of fundamental
importance to understanding many physical processes associated with the black
hole, as well as the assembly history of the SMBH population across cosmic
time. Over the past several decades, reverberation mapping \citep[RM,
e.g.,][]{Bahcall_etal_1972,Blandford_McKee_1982,Peterson_1993} has proven to
be a viable technique to measure (broad-line) AGN BH mass by providing an
estimate of the broad line region (BLR) size $R$
\citep[e.g.,][]{Peterson_etal_2004}, combined with the assumptions that the
BLR dynamics is dominated by the central BH mass and that the widths of the
broad emission lines $V$ are related to the virial velocity of the BLR
\citep[e.g.,][]{Dibai_1980,Wandel_etal_1999}. The unknown geometry of the BLR
is absorbed in a constant virial coefficient $f$, which is calibrated
\citep[e.g.,][]{Onken_etal_2004,Woo_etal_2010,Graham_etal_2011} to bring the
products of $RV^2/G$ into average agreement with those predicted from the
local scaling relation between BH mass and bulge velocity dispersion (the
$M-\sigma$ relation).

An important result of RM studies is the discovery of a tight correlation
between the BLR size and the continuum luminosity of broad-line AGNs
\citep[e.g.,][]{Kaspi_etal_2000,Bentz_etal_2006}, i.e., the $R-L$ relation,
when plotted over a wide dynamical range in AGN luminosity. This relation has
led to the development of the so-called single-epoch virial BH mass
estimators \citep[``virial BH mass estimators'' for short,
e.g.,][]{Vestergaard_2002,McLure_Jarvis_2002,Mclure_Dunlop_2004,Greene_Ho_2005,Vestergaard_Peterson_2006},
in which one measures the continuum (or line) luminosity and broad line width
from single-epoch spectroscopy to derive a virial product as the BH mass
estimate, with coefficients calibrated from a sample of $\sim 40$ local AGNs
with RM masses (which are further tied to the predictions from the $M-\sigma$
relation). Various versions of single-epoch virial BH mass estimators have
been developed since, based on different broad lines and advocating different
recipes for measuring luminosities and line widths
\citep[e.g.,][]{McGill_etal_2008,Wang_etal_2009b,Vestergaard_Osmer_2009,
Rafiee_Hall_2011a,Shen_etal_2011}. This empirical method, albeit rooted in
the RM technique, is much less expensive than RM, and hence has been applied
in numerous studies to estimate quasar/AGN BH masses, notably for large
statistical samples
\citep[e.g.,][]{Woo_Urry_2002,Mclure_Dunlop_2004,Kollmeier_etal_2006,Greene_Ho_2007a,
Vestergaard_etal_2008,Shen_etal_2008b,Shen_etal_2011}.

Despite the wide application of these virial BH mass estimators, there are
many statistical and systematic uncertainties of these estimates. First and
foremost, all single-epoch mass estimators are bootstrapped from a sample of
only $\sim 40$ $z\lesssim 0.4$ RM AGNs (consisting of Seyfert 1 galaxies and
several PG quasars), which is known to be unrepresentative of their
high-luminosity and high-redshift counterparts
\citep[e.g.,][]{Richards_etal_2011}. The statistics of RM AGNs need to be
substantially improved to account for the diversity in BLR properties.
Secondly, different versions of virial mass estimators have different
systematics depending on the quality of the spectrum and the profile of the
broad line, and there is currently no consensus as to which version is the
best. Nevertheless, there are some general considerations on various
estimators:

$\bullet$ {\em Which line to use:} The commonly utilized pairs of line and
luminosity in the restframe UV and optical are: \halpha\ with $L_{\rm
H\alpha}$ or $L_{5100}$, \hbeta\ with $L_{5100}$, \MgII\ with $L_{3000}$, and
\CIV\ with $L_{1350}$ (or $L_{1450}$). Since the Balmer lines \halpha\ and
\hbeta\ are the most studied lines in reverberation mapping and the $R-L$
relation was originally measured for the Balmer line BLR radius and
$L_{5100}$\citep[e.g.,][]{Kaspi_etal_2000,Bentz_etal_2006}, it is reasonable
to argue that the virial mass estimator based on the Balmer lines is the most
reliable one. The width of the broad \halpha\ is well correlated with that of
the broad \hbeta\ and therefore it provides a good substitution in the
absence of \hbeta\ \citep[e.g.,][]{Greene_Ho_2005}.

The \MgII\ line has not been studied much in RM \citep[cf.,][]{Woo_2008}, and
only in very few cases has a time-lag of \MgII\ been measured with RM
\citep[e.g.,][]{Reichert_etal_1994,Metzroth_etal_2006}; but the width of
\MgII\ is shown to correlate with that of \hbeta\ in single-epoch spectra
\citep[e.g.,][]{Salviander_etal_2007,Shen_etal_2008b,McGill_etal_2008,Wang_etal_2009b},
suggesting that \MgII\ may be used as a substitution for \hbeta\ in
estimating virial BH masses.

The \CIV\ line is known to vary and time-lags have been measured for \CIV\ in
several objects \citep[e.g.,][]{Peterson_etal_2004,Kaspi_etal_2007}, although
the sample is too small to derive a reliable $R-L$ relation for \CIV.
However, the high-ionization \CIV\ line differs from low-ionization lines
such as \MgII\ and the Balmer lines in many ways \citep[for a review, see
][]{Sulentic_etal_2000b}, most notably it shows a prominent blueshift with
respect to the low-ionization lines \citep[e.g.,][]{Gaskell_1982}. In
addition, the \CIV\ line is generally more asymmetric than \MgII\ and the
Balmer lines, and the width of \CIV\ is poorly correlated with those of
\MgII\ and \hbeta\
\citep[e.g.,][]{Baskin_Laor_2005,Netzer_etal_2007,Shen_etal_2008b}. The
different properties of \CIV\ suggest that \CIV\ is probably more affected by
a non-virial component such as arising from a radiatively-driven disk wind
\citep[e.g.,][]{Murray_etal_1995,Proga_etal_2000}, and would therefore be a
biased virial mass estimator \citep[e.g.,][and references
therein]{Baskin_Laor_2005,Sulentic_etal_2007,Netzer_etal_2007,Shen_etal_2008b,Marziani_Sulentic_2011}.
However, since both the Balmer lines and \MgII\ move out of the optical
bandpass at $z\gtrsim 2$, it would be useful to improve the \CIV\ estimator
in order to measure BH masses at high redshift without the need for near-IR
spectroscopy.

\citet{Shen_etal_2008b} used a large sample of $\sim 5000$ SDSS quasars to
show that the difference between the \CIV\ and \MgII\ virial masses is
correlated with the \CIV-\MgII\ blueshift. On the other hand,
\citet{Assef_etal_2011} used a sample of $\sim 10$ quasars with optical
spectra covering \CIV\ and near-IR spectra covering \hbeta/\halpha\ to show
that the difference between \CIV\ and Balmer line virial masses is largely
driven by their restframe UV-to-optical continuum luminosity ratio $L_{\rm
1350}/L_{5100}$, suggesting that much of the dispersion in their virial mass
difference is caused by the poor correlation between $L_{5100}$ and
$L_{1350}$ rather than between their line widths. A larger sample is needed
to test this result.

$\bullet$ {\em Line dispersion vs FWHM:} The two common choices of line width
are FWHM, and the second moment of the line (line dispersion, $\sigma_{\rm
line}$). Both FWHM and $\sigma_{\rm line}$ have advantages and disadvantages.
FWHM is easier to measure, less susceptible to noise in the wings and line
blending than $\sigma_{\rm line}$, but is more sensitive to the treatment of
the narrow line removal. Arguably $\sigma_{\rm line}$ is a better surrogate
for the virial velocity \citep[e.g.,][]{Collin_etal_2006}, although the
evidence is not very strong. Since currently all the RM BH masses are
computed using $\sigma_{\rm line,rms}$ measured from the rms spectra
\citep{Peterson_etal_2004}, ideally one would like to use $\sigma_{\rm
line}$, albeit not measured from rms spectra, in single-epoch virial mass
estimators. In practice, however, $\sigma_{\rm line}$ measured from
single-epoch spectra depends on the quality of the spectra, line profile, and
specific treatment of deblending, and could differ significantly from one
observation/analysis to another
\citep[e.g.,][]{Denney_etal_2009a,Fine_etal_2010,Rafiee_Hall_2011b,Assef_etal_2011}.
Therefore in terms of readiness and repeatability, $\sigma_{\rm line}$ is
less favorable than FWHM in single-epoch virial mass estimators. For these
reasons, we will not utilize line dispersion in the current study.

In this paper we investigate the reliability of the UV virial mass estimators
(in particular \CIV) compared with the \hbeta\ (or \halpha) estimator with a
carefully selected sample of quasars with good \CIV\ to \MgII\ coverage in
optical SDSS spectra and our own near-IR spectra covering \hbeta\ and
\halpha. Our sample probes the high-luminosity regime ($L_{5100}>10^{45.4}\,
{\rm erg\,s^{-1}}$, or $L_{\rm bol}\gtrsim 2.5\times 10^{46}\,{\rm
erg\,s^{-1}}$) of quasars, and thus such a study will provide confidence on
estimating virial BH masses for the most luminous quasars (such as $z\gtrsim
6$ quasars). Our sample is substantially larger than earlier samples in
similar studies, which enables us to draw more statistically significant
conclusions. We are interested in examining empirical correlations between
line widths and continuum luminosities of two different lines, and any
dependence of their virial mass difference on specific quasar properties. We
describe our sample and follow-up near-IR observations in \S\ref{sec:data}.
The procedure of measuring spectral properties is detailed in
\S\ref{sec:spec_measure} and the results are presented in \S\ref{sec:result}.
We discuss the results in \S\ref{sec:dis} and conclude in \S\ref{sec:con}.
Throughout this paper we adopt a flat $\Lambda$CDM cosmology with
$\Omega_{\Lambda}=0.7$, $\Omega_0=0.3$ and $H_0=70\,{\rm
km\,s^{-1}\,Mpc^{-1}}$.

\section{Data}\label{sec:data}

\subsection{Sample Selection}\label{subsec:sample}

We select our targets from the SDSS DR7 quasar catalog
\citep[e.g.,][]{Schneider_etal_2010,Shen_etal_2011} for follow-up near-IR
spectroscopy with the following two criteria:
\begin{enumerate}

\item[$\bullet$] redshift between $1.5$ and $2.2$ and avoiding redshift
    ranges where the \hbeta\ and \halpha\ lines fall in the telluric
    absorption bands in the near-infrared;

\item[$\bullet$] with good (S/N $>10$) SDSS spectra covering \CIV\
    through \MgII, and no broad absorption features or unusual continuum
    shapes.

\end{enumerate}

These criteria by design selects luminous quasars (bolometric luminosity
$L_{\rm bol}> {\rm a\ few}\times 10^{46}\,{\rm erg\,s^{-1}}$) as our targets,
but the resulting sample still covers a range of spectral diversities such as
the line width and velocity shift of each broad lines. In addition, host
contamination is generally negligible for these objects, which greatly
simplifies our model fits and interpretations. Our targets have a similar
color excess $\Delta(g-i)$ distribution \citep[where $\Delta(g-i)$ is the
deviation of $g-i$ color from the median $g-i$ color at each redshift;
see][]{Richards_etal_2003} as the underlying SDSS quasars in this redshift
range, but do not include any dust-reddened objects (defined as
$\Delta(g-i)\gtrsim 0.3$). $\sim 10\%$ (5/49) of the targets are radio-loud
\citep[$R\equiv f_{\rm \nu,\,6\,cm}/f_{\nu,\,2500\,\textrm{\AA}}>10$,
see][]{Shen_etal_2011} based on the {\em Faint Images of the Radio Sky at
Twenty-Centimeters} \citep[FIRST,][]{White_etal_1997} catalog, and the rest
11 targets are not in the FIRST footprint as of July 16, 2008.

\begin{figure*}[!h]
 \centering
 \includegraphics[width=0.9\textwidth]{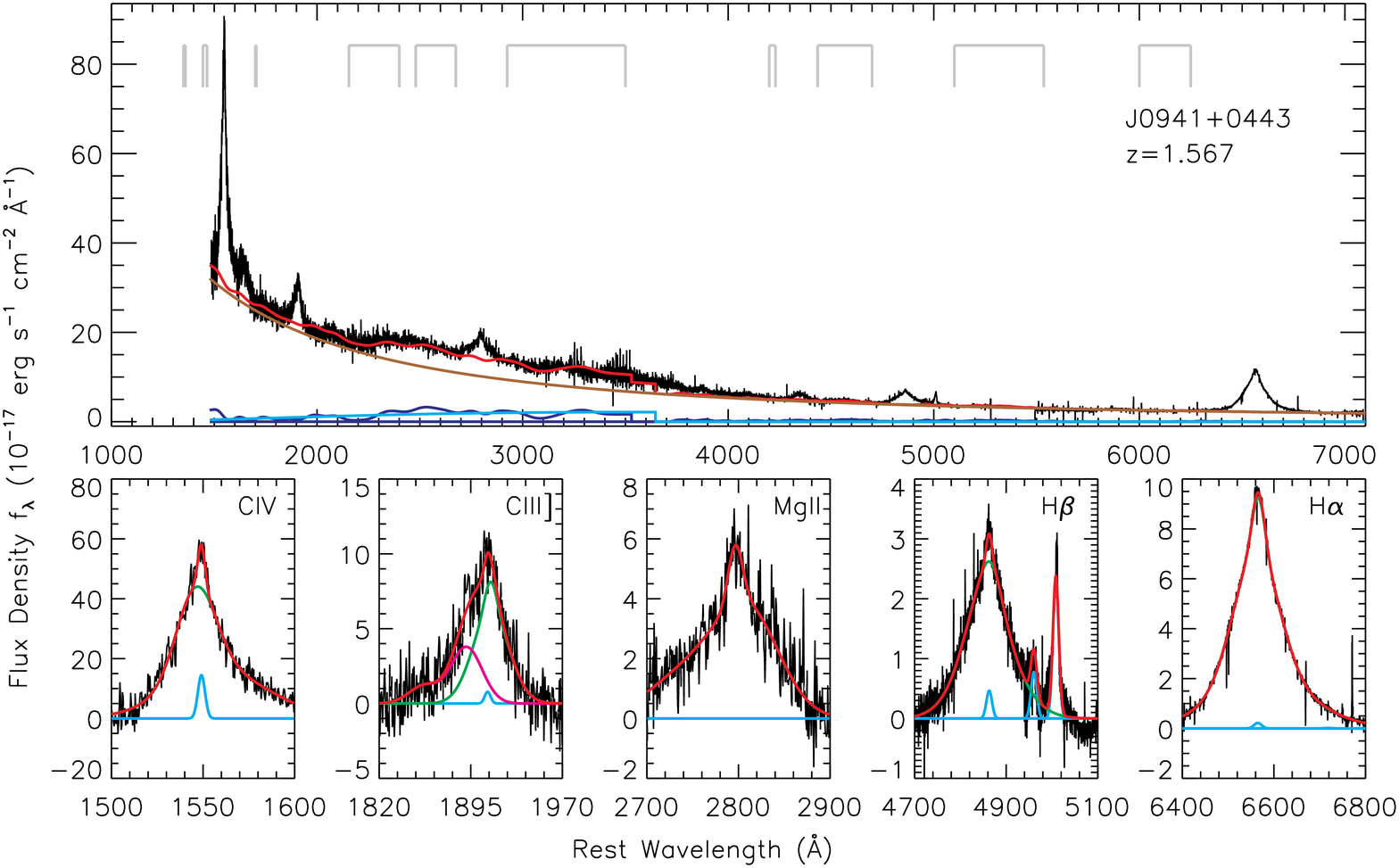}
    \caption{An example of our model fits to the combined optical and near-IR spectra (J0941$+$0443).
    The top panel shows the global fit of the pseudo-continuum, where the brown line
    is the power-law continuum, the blue line is the Fe II template fit, the cyan line
    is the Balmer continuum model, and the red line is the combined pseudo-continuum
    model to be subtracted off. The line segments near the top indicate the wavelength windows
    used for the pseudo-continuum fit. The bottom panels show the emission line fits to
    \CIV\ through \halpha, where the cyan lines are the model narrow line emission,
    the green lines are the model broad line emission, and the red lines are the combined
    model line profiles. For \CIII\ we also show the modeled \AlIII\ and \SiIII\ emission in
    magenta. }\label{fig:fit_examp}
\end{figure*}

\begin{deluxetable*}{ccccccccccccc}
\tablecaption{Sample Summary \label{table:sample}} \tablehead{ Object Name &
RA (J2000) & DEC (J2000) & Plate & Fiber & MJD & $z_{\rm HW}$ & $i_{\rm PSF}$ & $J_{\rm 2MASS}$ & $H_{\rm 2MASS}$ & $K_{\rm s,2MASS}$ & NIR Obs. &
Obs. UT\\
(1) & (2) & (3) & (4) & (5) & (6) & (7) & (8) & (9) & (10) & (11) & (12) &
(13)} \startdata
J0029$-$0956 &  00 29 48.04 & $-$09 56 39.4  & 0653 & 640 & 52145 & 1.618 & 17.672 & 16.728 & 15.747 & 15.622 & TSPEC & 100102/101128   \\
J0041$-$0947 &  00 41 49.64 & $-$09 47 05.0  & 0655 & 172 & 52162 & 1.629 & 16.966 & 16.201 & 15.680 & 15.535 & TSPEC & 100102/101128   \\
J0147$+$1332 &  01 47 05.42 & $+$13 32 10.0  & 0429 & 145 & 51820 & 1.595 & 17.115 & 16.194 & 15.473 & 15.474 & TSPEC & 090909/091107   \\
J0149$+$1501 &  01 49 44.43 & $+$15 01 06.6  & 0429 & 575 & 51820 & 2.073 & 17.275 & 16.565 & 15.998 & 15.243 & TSPEC & 090909/101128   \\
J0157$-$0048 &  01 57 33.87 & $-$00 48 24.4  & 0403 & 213 & 51871 & 1.551 & 18.164 & 16.797 & 16.553 &  0.000 & TSPEC & 091107/101128   \\
J0200$+$1223 &  02 00 44.50 & $+$12 23 19.1  & 0427 & 219 & 51900 & 1.654 & 17.811 & 16.573 & 16.125 &  0.000 & TSPEC & 100102/101128   \\
J0358$-$0540 &  03 58 56.73 & $-$05 40 23.4  & 0464 & 499 & 51908 & 1.506 & 18.258 & 17.572 & 16.297 &  0.000 & TSPEC & 100102/101128   \\
J0412$-$0612 &  04 12 55.16 & $-$06 12 10.3  & 0465 & 037 & 51910 & 1.691 & 17.322 & 16.306 & 16.077 & 15.353 & TSPEC & 100102/101128   \\
J0740$+$2814 &  07 40 29.82 & $+$28 14 58.5  & 0888 & 545 & 52339 & 1.545 & 17.445 & 16.426 & 15.689 & 15.482 & TSPEC & 091108   \\
J0812$+$0757 &  08 12 27.19 & $+$07 57 32.9  & 2570 & 026 & 54081 & 1.574 & 17.404 & 16.658 & 15.995 & 16.031 & TSPEC & 101202   \\
J0813$+$2545 &  08 13 31.28 & $+$25 45 03.0  & 1266 & 219 & 52709 & 1.513 & 15.385 & 14.085 & 13.271 & 13.056 & TSPEC & 091108   \\
J0813$+$1522 &  08 13 44.15 & $+$15 22 21.5  & 2270 & 439 & 53714 & 1.545 & 17.541 & 16.472 & 15.805 &  0.000 & TSPEC & 101122   \\
J0821$+$5712 &  08 21 46.22 & $+$57 12 26.0  & 1872 & 615 & 53386 & 1.546 & 16.868 & 15.943 & 15.027 & 15.031 & TSPEC & 091108/100104   \\
J0838$+$2611 &  08 38 50.15 & $+$26 11 05.4  & 1930 & 492 & 53347 & 1.618 & 16.098 & 15.211 & 14.424 & 14.288 & TSPEC & 091108  \\
J0844$+$2826 &  08 44 51.91 & $+$28 26 07.5  & 1588 & 179 & 52965 & 1.574 & 18.006 & 17.026 & 16.147 & 15.798 & TSPEC & 101202    \\
J0855$+$0029 &  08 55 43.26 & $+$00 29 08.5  & 0468 & 111 & 51912 & 1.525 & 17.952 & 16.829 & 16.545 & 15.668 & FIRE  & 110426     \\
J0917$+$0436 &  09 17 54.44 & $+$04 36 52.1  & 0991 & 284 & 52707 & 1.587 & 18.543 &  0.000 &  0.000 &  0.000 & FIRE  & 110427    \\
J0933$+$1413 &  09 33 18.49 & $+$14 13 40.1  & 2580 & 347 & 54092 & 1.561 & 17.465 & 16.520 & 15.540 & 15.448 & TSPEC & 100126    \\
J0941$+$0443 &  09 41 26.49 & $+$04 43 28.7  & 0570 & 379 & 52266 & 1.567 & 17.824 & 16.954 & 16.084 & 15.706 & FIRE  & 110427    \\
J0949$+$1751 &  09 49 13.05 & $+$17 51 55.9  & 2370 & 184 & 53764 & 1.675 & 17.143 & 16.137 & 15.626 & 15.332 & TSPEC & 100126    \\
J1004$+$4231 &  10 04 01.27 & $+$42 31 23.1  & 1217 & 573 & 52672 & 1.666 & 16.764 & 15.795 & 15.376 & 15.080 & TSPEC & 100104    \\
J1009$+$0230 &  10 09 30.51 & $+$02 30 52.4  & 0502 & 429 & 51957 & 1.557 & 18.556 & 17.310 &  0.000 &  0.000 & FIRE  & 110426   \\
J1014$+$5213 &  10 14 47.54 & $+$52 13 20.2  & 0904 & 259 & 52381 & 1.552 & 17.334 & 16.705 & 15.929 & 15.752 & TSPEC & 110124    \\
J1015$+$1230 &  10 15 04.75 & $+$12 30 22.2  & 1745 & 148 & 53061 & 1.703 & 17.400 & 16.374 & 15.989 & 15.658 & TSPEC & 110124    \\
J1046$+$1128 &  10 46 03.22 & $+$11 28 28.1  & 1601 & 193 & 53115 & 1.607 & 17.784 &  0.000 &  0.000 &  0.000 & FIRE  & 110426    \\
J1049$+$1432 &  10 49 10.31 & $+$14 32 27.1  & 1749 & 571 & 53357 & 1.540 & 17.740 & 16.813 & 15.590 & 15.458 & TSPEC & 100126   \\
J1059$+$0909 &  10 59 51.05 & $+$09 09 05.7  & 1220 & 231 & 52723 & 1.690 & 16.771 & 15.620 & 15.094 & 14.411 & TSPEC & 110124   \\
J1102$+$3947 &  11 02 40.16 & $+$39 47 30.1  & 1437 & 205 & 53046 & 1.664 & 17.605 & 16.563 & 16.153 & 15.811 & TSPEC & 110222   \\
J1119$+$2332 &  11 19 49.30 & $+$23 32 49.1  & 2493 & 077 & 54115 & 1.626 & 17.338 & 16.230 & 15.465 & 15.322 & TSPEC & 110124   \\
J1125$+$0001 &  11 25 42.29 & $+$00 01 01.3  & 0280 & 077 & 51612 & 1.692 & 17.305 & 16.503 & 15.573 & 15.137 & FIRE  & 110427   \\
J1138$+$0401 &  11 38 29.33 & $+$04 01 01.0  & 0838 & 241 & 52378 & 1.567 & 16.887 & 16.064 & 15.169 & 15.426 & FIRE  & 110426   \\
J1140$+$3016 &  11 40 23.40 & $+$30 16 51.5  & 2220 & 577 & 53795 & 1.599 & 16.680 & 15.827 & 14.903 & 14.989 & TSPEC & 100126   \\
J1220$+$0004 &  12 20 39.45 & $+$00 04 27.6  & 0288 & 516 & 52000 & 2.048 & 17.200 & 16.337 & 15.992 & 15.102 & FIRE  & 110427   \\
J1233$+$0313 &  12 33 55.21 & $+$03 13 27.6  & 0520 & 536 & 52288 & 1.528 & 17.814 & 16.745 & 16.294 &  0.000 & FIRE  & 110427   \\
J1234$+$0521 &  12 34 42.16 & $+$05 21 26.7  & 0846 & 341 & 52407 & 1.550 & 16.992 & 16.372 & 15.448 & 15.408 & TSPEC & 110513   \\
J1240$+$4740 &  12 40 06.70 & $+$47 40 03.3  & 1455 & 424 & 53089 & 1.561 & 17.507 & 16.573 & 15.791 &  0.000 & TSPEC & 110222   \\
J1251$+$0807 &  12 51 40.82 & $+$08 07 18.4  & 1792 & 427 & 54270 & 1.607 & 16.907 & 15.975 & 15.068 & 14.717 & FIRE  & 110426    \\
J1333$+$0058 &  13 33 21.90 & $+$00 58 24.3  & 0298 & 455 & 51955 & 1.511 & 17.776 & 16.888 & 16.039 & 15.683 & FIRE  & 110426  \\
J1350$+$2652 &  13 50 23.68 & $+$26 52 43.1  & 2114 & 105 & 53848 & 1.624 & 17.042 & 16.110 & 15.490 & 15.548 & TSPEC & 110222  \\
J1354$+$3016 &  13 54 39.70 & $+$30 16 49.2  & 2116 & 486 & 53854 & 1.553 & 17.680 & 17.137 & 15.574 & 15.844 & TSPEC & 110422  \\
J1419$+$0606 &  14 19 49.39 & $+$06 06 54.0  & 1826 & 183 & 53499 & 1.649 & 17.176 & 16.661 & 15.935 &  0.000 & FIRE  & 110426   \\
J1421$+$2241 &  14 21 08.71 & $+$22 41 17.4  & 2786 & 589 & 54540 & 2.188 & 16.906 & 15.632 & 14.962 & 14.019 & TSPEC & 100520/110513   \\
J1428$+$5925 &  14 28 41.97 & $+$59 25 52.0  & 0789 & 591 & 52342 & 1.660 & 17.418 & 16.800 & 15.803 & 15.461 & TSPEC & 110414/110418   \\
J1431$+$0535 &  14 31 48.09 & $+$05 35 58.0  & 1828 & 300 & 53504 & 2.095 & 16.523 & 15.368 & 14.892 & 14.166 & TSPEC & 100520  \\
J1432$+$0124 &  14 32 30.57 & $+$01 24 35.1  & 0535 & 054 & 51999 & 1.542 & 17.640 & 16.406 & 15.900 & 15.525 & FIRE  & 110427  \\
J1436$+$6336 &  14 36 45.80 & $+$63 36 37.9  & 2947 & 444 & 54533 & 2.066 & 16.528 & 15.443 & 15.014 & 14.201 & TSPEC & 100520/110513   \\
J1521$+$4705 &  15 21 11.86 & $+$47 05 39.1  & 1331 & 256 & 52766 & 1.517 & 17.531 & 16.668 & 15.836 &  0.000 & TSPEC & 110422  \\
J1538$+$0537 &  15 38 59.45 & $+$05 37 05.3  & 1836 & 377 & 54567 & 1.684 & 17.889 & 16.905 & 16.179 &  0.000 & FIRE  & 110426  \\
J1542$+$1112 &  15 42 12.90 & $+$11 12 26.7  & 2516 & 165 & 54240 & 1.540 & 17.295 & 17.083 & 15.702 &  0.000 & FIRE  & 110427   \\
J1552$+$1948 &  15 52 40.40 & $+$19 48 16.7  & 2172 & 390 & 54230 & 1.613 & 17.450 & 16.547 & 15.903 & 15.746 & TSPEC & 110414/110418   \\
J1604$-$0019 &  16 04 56.14 & $-$00 19 07.1  & 0344 & 155 & 51693 & 1.636 & 17.072 & 16.219 & 15.281 & 15.420 & FIRE  & 110426  \\
J1621$+$0029 &  16 21 03.98 & $+$00 29 05.8  & 0364 & 353 & 52000 & 1.689 & 18.489 & 17.255 &  0.000 &  0.000 & FIRE  & 110714   \\
J1710$+$6023 &  17 10 30.20 & $+$60 23 47.5  & 0351 & 004 & 51780 & 1.549 & 17.359 & 16.446 & 15.426 & 15.092 & TSPEC & 110414/110418   \\
J2040$-$0654 &  20 40 09.62 & $-$06 54 02.5  & 0634 & 088 & 52164 & 1.611 & 18.850 &  0.000 &  0.000 &  0.000 & FIRE  & 110426 \\
J2045$-$0101 &  20 45 36.56 & $-$01 01 47.9  & 0982 & 278 & 52466 & 1.661 & 16.415 & 15.650 & 14.889 & 14.672 & FIRE  & 110427 \\
J2045$-$0051 &  20 45 38.96 & $-$00 51 15.5  & 0982 & 277 & 52466 & 1.590 & 18.063 & 17.047 &  0.000 &  0.000 & FIRE  & 110426  \\
J2055$+$0043 &  20 55 54.08 & $+$00 43 11.4  & 0984 & 326 & 52442 & 1.624 & 18.594 & 17.308 &  0.000 &  0.000 & FIRE  & 110427  \\
J2137$+$0012 &  21 37 48.44 & $+$00 12 20.0  & 0989 & 585 & 52468 & 1.670 & 18.046 & 16.864 & 16.096 & 16.042 & FIRE  & 110713  \\
J2232$+$1347 &  22 32 46.80 & $+$13 47 02.0  & 0738 & 520 & 52521 & 1.557 & 17.340 & 16.208 & 15.531 & 15.372 & TSPEC & 091107   \\
J2258$-$0841 &  22 58 00.02 & $-$08 41 43.7  & 0724 & 571 & 52254 & 1.496 & 17.459 & 16.893 & 16.288 &  0.000 & TSPEC & 091107  \\
\enddata
\tablecomments{Summary of the sample of SDSS quasars for which we have
conducted near-infrared spectroscopy. Columns (4)-(6): plate, fiber and MJD
of the optical SDSS spectrum for each object; (7): improved quasar redshift
from \citet{Hewett_Wild_2010}; (8): SDSS $i$-band PSF magnitudes; (9)-(11):
2MASS (Vega) magnitudes; (12): instrument for the near-IR spectroscopy; (13):
UT dates of the near-IR observations. Note that here the 2MASS magnitudes
were taken from \citet{Schneider_etal_2010}, where aperture photometry was
performed upon 2MASS images to detect faint objects, hence these near
infrared data go beyond the 2MASS All-Sky and ``$6\times$'' point source
catalogs \citep[see][ for details]{Schneider_etal_2010}. }
\end{deluxetable*}

\subsection{Near-IR Spectroscopy}\label{subsec:nir}

We observed our targets during 2009-2011 with TripleSpec
\citep{Wilson_etal_2004} on the ARC 3.5\,m telescope, and with the
Folded-port InfraRed Echellette \citep[FIRE,][]{Simcoe_etal_2010} on the
6.5\,m Magellan-Baade telescope. Table \ref{table:sample} summarizes our
sample and follow-up observations. Below we describe the observations and
data reduction for TripleSpec and FIRE data, respectively.

\subsubsection{ARC 3.5m/TripleSpec}\label{subsec:triplespec}

TripleSpec \citep{Wilson_etal_2004} is a near-IR spectrograph with
simultaneous $0.95-2.46$\,\micron\ overage. We observed our targets during
2009-2011 semesters. The total exposure time varied from object to object due
to different target brightness and observing conditions, but is typically
$1-1.5$\,hr. We used slits with widths of both 1.1\arcsec\ and 1.5\arcsec\
during the course of the observations, and the resulting spectral resolution
is $R\sim 2500-3500$. The slit was positioned at the parallactic angle in the
middle of the observation, and we performed standard ABBA dither patterns to
aid sky subtraction. For each object we observed a nearby A0V star as flux
and telluric standard immediately before or after observing the science
target.

We reduced the Triplespec data using the IDL-based pipeline
\texttt{APOTripleSpecTool}, which is a modified version of the Spextool
package developed by Michael Cushing \citep{Cushing_etal_2004}. The reduction
procedures include non-linearity correction, flat-fielding, wavelength
calibration using OH sky lines (calibrated to vacuum wavelength), sky
subtraction using adjacent exposures at nodding slit positions, cosmic-ray
rejection, optimal extraction of 1-D spectra \citep{Horne_1986}, combining
individual exposures, merging multiple echelle orders, and heliocentric
corrections. We used the A0V-star observations for relative flux calibration
and telluric correction following the technique of \citet{Vacca_etal_2003}
using the \texttt{xtellcor} routine contained in the Spextool package
(Cushing et al. 2004). We tied the absolute flux calibration to the Two
Micron All Sky Survey (2MASS) \citep{Skrutskie_etal_2006} $H$-band magnitudes
using synthetic magnitudes computed from our spectrum with the 2MASS relative
spectral response curves in \citet{Cohen_etal_2003}. This absolute flux
calibration neglects the continuum variability between the 2MASS and
(spectroscopic) SDSS epochs, which is typically at the level of $\sim
0.1$\,mag for average SDSS quasars
\citep[e.g.,][]{Sesar_etal_2007,MacLeod_etal_2011}. It also neglects possible
line shape variability of quasars between the two epochs of SDSS and near-IR
observations, but this variation is likely negligible ($\sigma_{\rm
FWHM}<0.05$ dex) based on repeated spectroscopy of the same objects
\citep[e.g.,][]{Wilhite_etal_2007,Park_etal_2011}.

Finally, for 14 targets we have a second observation on a different night. We
combined these repeated observations using the inverse-variance weighted mean
of the two observations.

\subsubsection{Magellan/FIRE}\label{subsec:fire}

FIRE \citep[][]{Simcoe_etal_2010} is a near-IR echelle spectrometer covering
the full 0.8-2.5\,\micron\ band. We observed 20 targets during the nights of
April 25-26, 2011, and another two targets on the nights of July 12-13, 2011.
We used the 0.6\arcsec\ slit width in Echelle mode, which offers a spectral
resolution of $R\sim 6000$ (50\,${\rm km\,s^{-1}}$). Typical total exposure
times were 45\,min per target but varied from object to object. We observed
our targets at the parallactic angle, and for each target we observed a
nearby A0V star for flux and telluric standard.

We reduced the FIRE data using the IDL-based pipeline ``FIREHOSE'' developed
by Robert Simcoe et al
\footnote{http://web.mit.edu/$\sim$rsimcoe/www/FIRE/ob\_data.htm}. The
reduction procedures are similar to those for the TripleSpec data with the
exception of sky subtraction. Instead of subtracting adjacent nodding
exposures, sky subtraction was performed using a B-spline model of the sky
directly constructed for each exposure following the technique of
\citet{Kelson_2003}.

The FIRE spectra have a substantial spectral overlap with the SDSS spectra.
Therefore we used the common part with the SDSS spectrum to normalize the
FIRE spectral flux density. As for our TripleSpec data, we neglect variations
in line shape between the SDSS and FIRE spectroscopic epoches.

\section{Spectral Measurements}\label{sec:spec_measure}


To derive line width and continuum luminosities used in single-epoch virial
mass estimators, we perform spectral fits to the optical and near-IR spectra,
as commonly adopted in the literature
\citep[e.g.,][]{Greene_Ho_2005,Salviander_etal_2007,Shen_etal_2008b,Wang_etal_2009b}.
Spectral fits with some functional form have certain advantage of being less
susceptible to noise than direct spectral measurements, although sometimes
there are still ambiguities in decomposing the spectrum into different
components.

Here we perform least-$\chi^2$ global fits to the combined optical and
near-IR spectra for the same object. Such global fits were not possible for
objects with limited wavelength coverage \citep[e.g.,][]{Shen_etal_2011}.
Each combined spectrum was de-reddened for Galactic extinction using the
\citet{CCM_1989} Milky Way reddening law and $E(B-V)$ derived from the
\citet{SFD_1998} dust map. The spectrum was then shifted to restframe using
the improved redshifts provided by \citet{Hewett_Wild_2010} for SDSS quasars,
where the spectral fit was performed. For each object we masked out narrow
absorption line features imprinted on the spectrum, which will bias the
continuum and emission line fits.

\subsection{Pseudo-Continuum Fit}

We first fit a pseudo-continuum model to account for the power-law (PL)
continuum, Fe II emission and Balmer continuum underneath the broad emission
lines of interest. All components were fit simultaneously. Templates for Fe
II and Fe III emission have been constructed from the spectrum of the
narrow-line Seyfert 1 galaxy, I Zw 1
\citep[e.g.,][]{Boroson_Green_1992,Vestergaard_Wilkes_2001,Tsuzuki_etal_2006}.
In this work we do not include additional Fe III emission in the fits as we
found this component is poorly constrained \citep[e.g.,][]{Greene_etal_2010}.
For the UV Fe II template, we use the \citet{Vestergaard_Wilkes_2001}
template (1000-3090\,\AA). \citet{Salviander_etal_2007} modified this
template by extrapolating below the \MgII\ line, and we use their template
for the 2200-3090\,\AA\ region; we augment the 3090-3500\,\AA\ region using
the template derived by \citet{Tsuzuki_etal_2006}. For the optical Fe II
template (3686-7484\,\AA) we use the one provided by
\citet{Boroson_Green_1992}. The PL continuum model has two free parameters,
the normalization and the PL slope. The UV and optical Fe II templates are
fitted independently, each has three free parameters, the normalization
factor, the velocity dispersion (to be convolved with the template), and the
wavelength shift of the template. The Fe II templates are only used as an
approximation to remove significant iron emission, and we found that they did
a reasonably good job. However, we are not concerned with the properties of
the iron emission in this work, and thus we do not interpret the physical
meanings of the velocity dispersion and wavelength shift of the Fe II
templates.

For the Balmer continuum we follow the empirical model by \citet{Grandi_1982}
as composed of partially optically thick clouds with an effective temperature
\citep[e.g.,][]{Dietrich_etal_2002,Wang_etal_2009b,Greene_etal_2010}:
\begin{equation}
f_{\rm BC}(\lambda) = AB_{\lambda}(\lambda,T_e)(1-e^{-\tau_\lambda});\quad \lambda\le \lambda_{\rm BE}\,
\end{equation}
where $T_e$ is the effective temperature, $\lambda_{\rm BE}\equiv 3646\,$\AA\
is the Balmer edge, $\tau_{\lambda}=\tau_{\rm BE}(\lambda/\lambda_{\rm
BE})^3$ is the optical depth with $\tau_{\rm BE}$ the optical depth at
$\lambda_{\rm BE}$, $A$ is the normalization factor, and
$B_{\lambda}(\lambda,T_e)$ is the Planck function at temperature $T_e$.
During the continuum fits, we have three free parameters, $1\times
10^4<T_e<5\times 10^4\,$K, $0.1<\tau_{\rm BE}<2$, and $A>0$.

Note that for limited wavelength fitting range (i.e., $\lambda<3646\,$\AA),
the Balmer continuum cannot be well constrained and is degenerate with the
power-law and Fe II components \citep[e.g.,][]{Wang_etal_2009b}, and is
generally not fitted \citep[e.g.,][]{Shen_etal_2011}. In the case of global
fits, a single power-law continuum is required to simultaneously fit the
region from \CIV\ to \halpha, providing some additional constraints on the
Balmer continuum; however even in this case, the Balmer continuum may still
be poorly constrained in a few cases, which will lead to uncertainties in the
power-law continuum luminosity estimates. Nevertheless, the isolation of
broad emission lines is not affected much by including or excluding the
Balmer continuum model.

We fit the pseudo-continuum model to a set of continuum windows free of
strong emission lines (except for Fe II): 1350-1360\,\AA, 1445-1465\,\AA,
1700-1705\,\AA, 2155-2400\,\AA, 2480-2675\,\AA, 2925-3500\,\AA,
4200-4230\,\AA, 4435-4700\,\AA, 5100-5535\,\AA, 6000-6250\,\AA,
6800-7000\,\AA. We try fitting both with and without the Balmer continuum
component and adopt the fit with the lower reduced $\chi^2$ value; usually
adding the Balmer continuum improves the global fit.

In a few cases ($\sim 5$ objects) we found that this global pseudo-continuum
model does not fit the \CIV-\CIII\ region well, which is likely caused by
intrinsic reddening in these systems, ill-determined Balmer continuum
strength, or mis-matched iron template. For these objects we perform local
($\lambda<2165$\,\AA\ with the same continuum windows defined above)
continuum fits around the \CIV\ and \CIII\ regions without the Balmer
continuum and Fe II emission (i.e., only with the PL component), in order to
get better measurements for \CIV\ and \CIII. The PL continuum model is then
used to estimate the monochromatic continuum luminosity $L_\lambda=\lambda
f_\lambda$ at 5100\,\AA, 3000\,\AA\ and 1350\,\AA. Although some of our
targets do not have spectral coverage of the restframe 1350\,\AA\ due to
their relatively lower redshift, the global PL component is well constrained
so the extrapolation of the model continuum to 1350\,\AA\ is not a problem.

\subsection{Emission Line Fits}

Once we have constructed the pseudo-continuum model, we subtract it from the
original spectrum, leaving the emission-line spectrum. We then fit the
\halpha, \hbeta, \MgII, \CIII, \CIV\ broad line complexes simultaneously with
mixtures of Gaussians (in logarithmic wavelength), as detailed below:

\begin{enumerate}

\item[$\bullet$]\halpha: we fit the wavelength range 6400-6800\,\AA. We
    use up to 3 Gaussians for the broad \halpha\ component, 1 Gaussian
    for the narrow \halpha\ component, 2 Gaussians for the \NIIa\ and
    \NIIb\ narrow lines, and 2 Gaussians for the \SIIa\ and \SIIb\ narrow
    lines. Since the narrow \NII\ lines are underneath the broad \halpha\
    profile, we tie their flux ratio to be $f_{6584}/f_{6548}=3$ to
    reduce ambiguities in decomposing the \halpha\ complex.

\item[$\bullet$]\hbeta: we fit the wavelength range 4700-5100\,\AA. We
    use up to 3 Gaussians for the broad \hbeta\ component and 1 Gaussian
    for the narrow \hbeta\ component. We use 2 Gaussians for the \OIIIa\
    and \OIIIb\ narrow lines. Given the quality of the near-IR spectra,
    we decided to only fit single Gaussians to the \OIIIab\ lines, and we
    tie the flux ratio of the \OIII\ doublet to be $f_{5007}/f_{4959}=3$.

\item[$\bullet$]\MgII: we fit the wavelength range 2700-2900\,\AA. We use
    up to 3 Gaussians for the broad \MgII\ component and 1 Gaussian for
    the narrow \MgII\ component. We do not try to fit the \MgII\ lines as
    a doublet, as the line splitting is small enough not to affect the
    broad line width measurements, and the spectral quality is usually
    inadequate for fitting a doublet to the narrow \MgII\ emission.

\item[$\bullet$]\CIII: we fit the wavelength range 1820-1970\,\AA. We use
    up to 2 Gaussians for the broad \CIII\ component and 1 Gaussian for
    the narrow \CIII\ component. We use two additional Gaussians for the
    \SiIIIwave\ and \AlIIIwave\ lines adjacent to \CIII. To reduce
    ambiguities in decomposing the \CIII\ complex, we tie the centroids
    of the two Gaussians for the broad \CIII\, i.e., the broad \CIII\
    profile is forced to be symmetric; we also tie the velocity offsets
    of \SiIII\ and \AlIII\ to their relative laboratory velocity offset.

\item[$\bullet$]\CIV: we fit the wavelength range 1500-1600\,\AA. We use
    up to 3 Gaussians for the broad \CIV\ component and 1 Gaussian for
    the narrow \CIV\ component. We do not fit the 1640\,\AA\ \HeII\
    feature as its contribution blueward of 1600\,\AA\ is negligible and
    will not bias the \CIV\ line fit.

\item[$\bullet$] During the line fitting, all narrow line components are
    constrained to have the same velocity offset and line width. We also
    impose an upper limit of $1200\,{\rm km\,s^{-1}}$ for the FWHM of the
    narrow line component\footnote{This upper limit is slightly larger
    than the values used in some studies (typically $\sim 750-1000\,{\rm
    km\,s^{-1}}$). For luminous SDSS quasars, \OIII\ FWHM values
    exceeding $\sim 1000\,{\rm km\,s^{-1}}$ are often seen
    \citep[e.g.,][]{Shen_etal_2011}. We hereby adopt the $1200\,{\rm
    km\,s^{-1}}$ upper limit for the narrow line width.}.

\end{enumerate}

While the presence of narrow line components for the Balmer lines is beyond
doubt, the relative contribution from narrow line components for the UV lines
is less certain. For \MgII, there is clear evidence that a narrow line
component is present at least in some quasars
\citep[e.g.,][]{Shen_etal_2008b,Shen_etal_2011,Wang_etal_2009b}. For \CIV,
the \citet{Vestergaard_Peterson_2006} virial mass calibration uses the FWHM
from the whole line profile, while some argue that a narrow line component
should be subtracted for \CIV\ as well \citep[e.g.,][]{Baskin_Laor_2005}. The
presence of narrow emission lines in the restframe optical spectra is
essential to provide constraints on the narrow line contribution for the UV
lines. We will measure the \CIV\ line width both with and without narrow line
subtraction and test if it is necessary to remove narrow line emission for
\CIV.

\subsection{Measurement Uncertainties}

It is important to quantify the uncertainties in our spectral measurements.
The nature of the non-linear model and multi-component fits introduces
ambiguities in decomposition, and the resulting uncertainties are usually
larger than those estimated from the parameter co-variance matrix of the
least-$\chi^2$ fits. We estimate the uncertainties in measured spectral
quantities using a Monte-Carlo approach as in \citet{Shen_etal_2011}. For
each object we generate 50 random realizations of mock spectra by adding
Gaussian noise to the original spectrum at each pixel using the spectral
error array. Technically speaking, these mock spectra are not the exact
alternative realizations of the original spectrum since the errors were added
twice, but they are only slightly degraded realizations and provide a good
approximation to capture the wavelength-dependent noise properties. We fit
each mock spectrum with the same fitting procedure described above and derive
the distribution of each measured spectral quantity (such as FWHM, velocity
offset, etc). We then take the semi-quartile of the 68$\%$ range of the
distribution as the nominal uncertainty of the measured quantity. This
approach takes into account the statistical uncertainties due to flux errors,
and systematic uncertainties due to ambiguities in decomposing multiple
components.

Fig.\ \ref{fig:fit_examp} shows an example of our global fits, and we
tabulate the measured quantities in Table \ref{table:measure}. Although with
different fitting recipes, the continuum and emission line measurements are
consistent with the measurements with SDSS spectra alone
\citep{Shen_etal_2011}, and the largest discrepancy occurs for $L_{\rm
MgII,broad}$ and $L_{3000}$: $\log L_{3000}$ is systematically smaller by
$\sim 0.12$ dex, and $\log L_{\rm MgII,broad}$ is systematically larger by
$\sim 0.067$ dex, than the measurements in \citet{Shen_etal_2011}. This is
largely caused by the additional Balmer continuum model in the spectral fits.
For simplicity, from now on we will by default refer to the broad line
component when we mention the FWHM or luminosity of a particular line unless
stated otherwise.

\section{Results}\label{sec:result}

We now proceed to examine correlations between continuum (line) luminosities
and between line widths for different lines, as well as their virial
products.

\subsection{Luminosity Correlations}\label{subsec:corr_lum}

\begin{figure}
 \centering
 \includegraphics[width=0.48\textwidth]{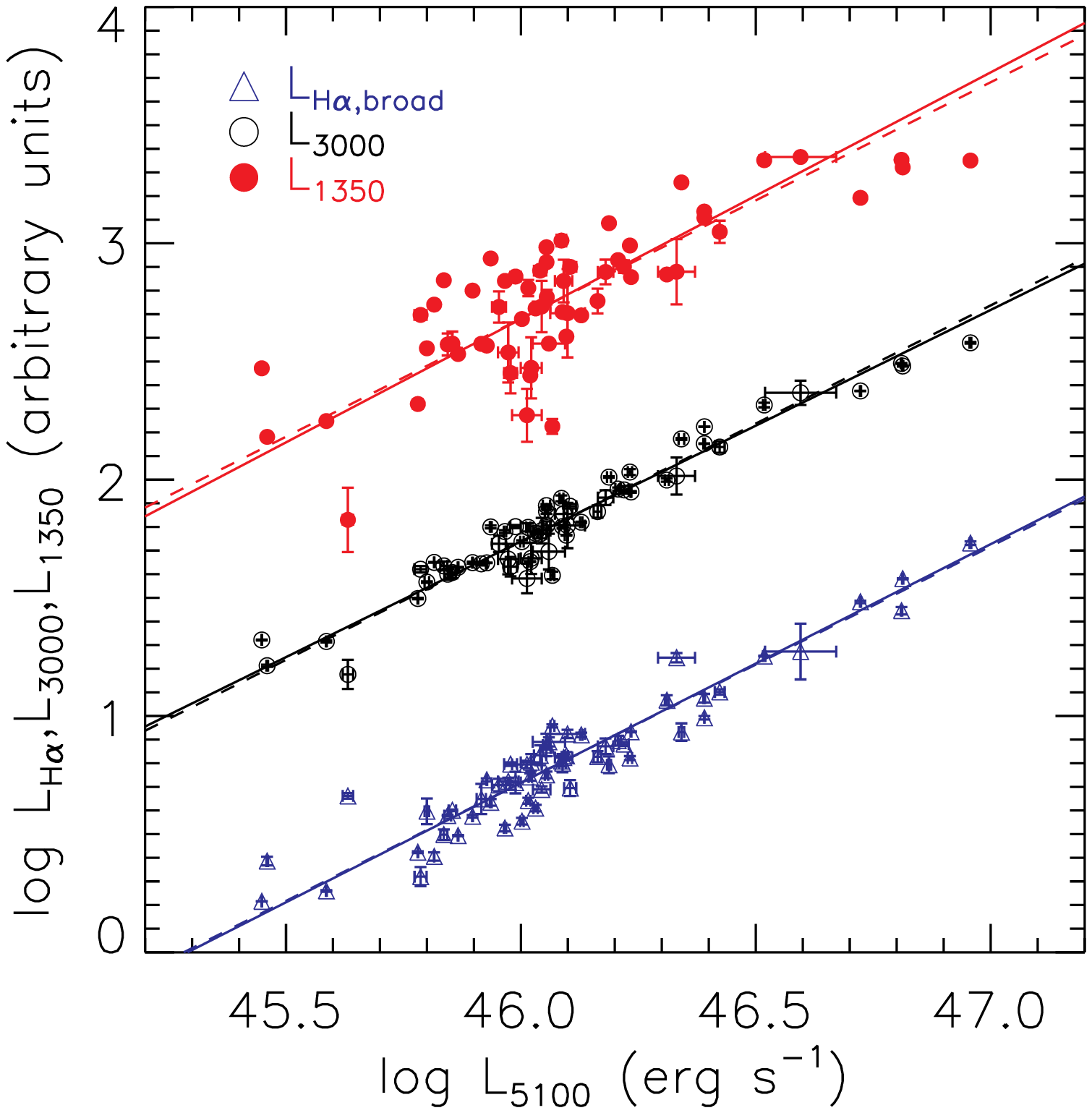}
 \includegraphics[width=0.48\textwidth]{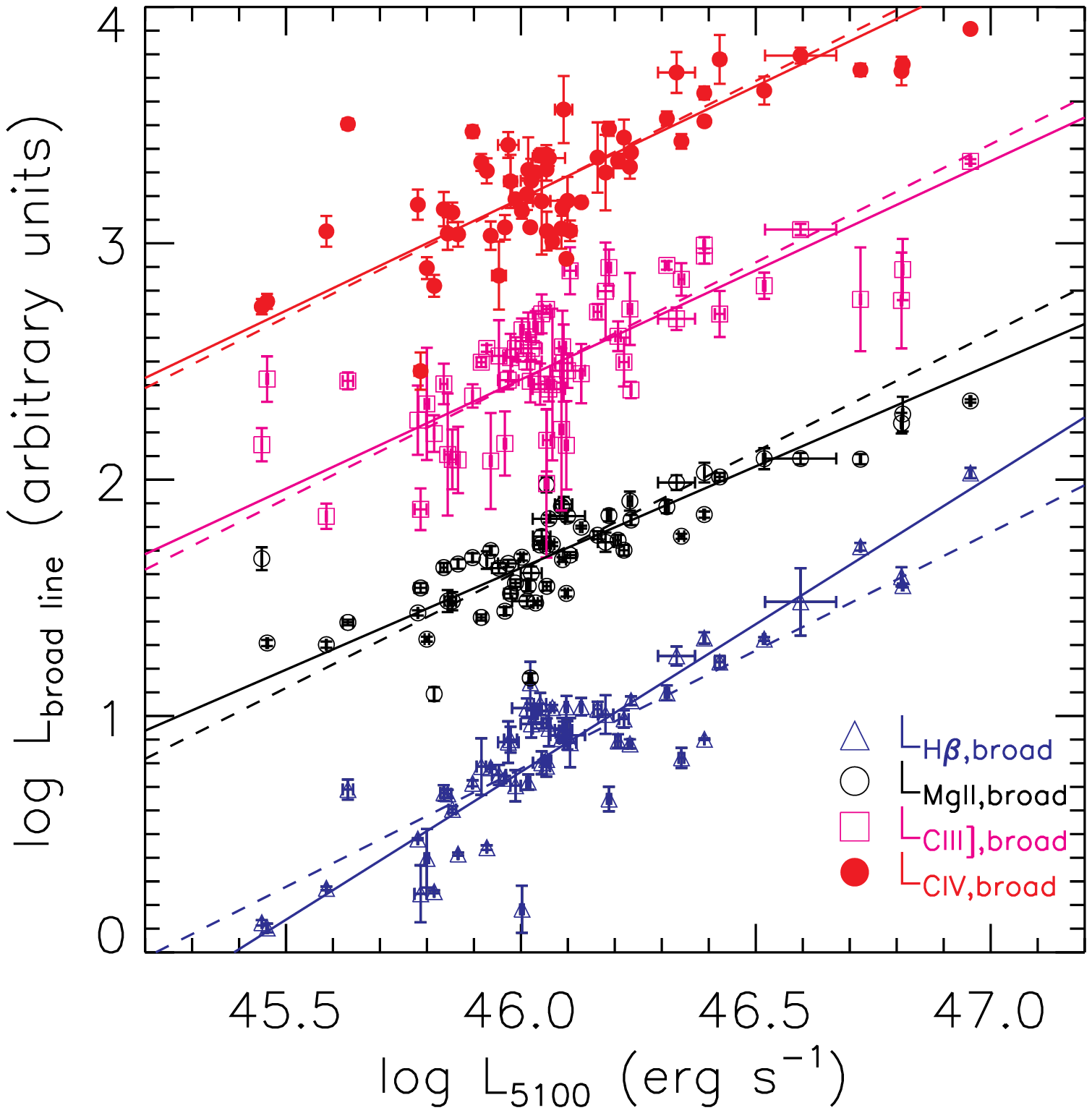}
    \caption{Correlations of different luminosity indicators with $L_{5100}$ for our sample. Each correlation
    has been shifted vertically for clarity without changing the scatter in the correlation.
    {\em Upper:} Correlations between $L_{5100}$ and the three most-frequently
    used alternative luminosity indicators for the BLR size, $L_{3000}$, $L_{1350}$ and
    $L_{\rm H\alpha,broad}$. {\em Bottom:}
    Correlations between $L_{5100}$ and broad line luminosities. In both panels the solid lines
    are the bisector linear regression results using the BCES estimator \citep[e.g.,][]{Akritas_Bershady_1996},
     and the dashed lines indicate a linear correlation of unity slope. }\label{fig:lum}
\end{figure}

In Fig.\ \ref{fig:lum} we compare different luminosities with the continuum
luminosity at 5100\,\AA, and we list the slopes from the bisector linear
regression fits using the BCES estimator \citep{Akritas_Bershady_1996} in
Table \ref{table:bces}. Our objects all have luminosity
$L_{5100}>10^{45.4}\,{\rm erg\,s^{-1}}$, and therefore contamination from
host starlight is mostly negligible \citep[e.g.,][]{Shen_etal_2011}. For UV
estimators, $L_{3000}$ and $L_{1350}$ (or $L_{1450}$) are often used in
replacement of $L_{5100}$. In addition, the luminosity of the broad \halpha\
line $L_{\rm H\alpha}$ is also used in pair with \halpha\ line width
\citep[e.g.,][]{Greene_Ho_2005}. Since the original $R-L$ relation is
calibrated against $L_{5100}$, a good correlation with $L_{5100}$ is required
to produce a reasonable estimate of the BLR size with alternative luminosity
indicators. As shown in the top panel of Fig.\ \ref{fig:lum}, all three
luminosity indicators are correlated with $L_{5100}$, where $L_{H\alpha}$ and
$L_{3000}$ are better correlated with $L_{5100}$ than $L_{1350}$\footnote{The
measured $L_{3000}$ with the Balmer continuum component in the fit is on
average smaller by $\sim 0.12$ dex than that without fitting the Balmer
continuum. However, both measurements of $L_{3000}$ are tightly correlated
with $L_{5100}$.}. The slopes of these correlations are very close to unity,
which means the ratios of these luminosities to $L_{5100}$ almost do not
depends on luminosity over this luminosity range.

In cases where the continuum is too faint to detect or contaminated by host
starlight or emission from a relativistic jet, an alternative route is to use
the luminosity of the broad lines
\citep[e.g.,][]{Wu_etal_2004,Greene_Ho_2005,Shen_etal_2011}. The bottom panel
of Fig.\ \ref{fig:lum} shows correlations of different line luminosities with
$L_{5100}$. Again, the \hbeta\ and \MgII\ line luminosities seem to correlate
with $L_{5100}$ with lower scatter than \CIII\ and \CIV. Interestingly, the
scatter in the $L_{\rm CIV}-L_{5100}$ relation is comparable to that in the
$L_{\rm 1350}-L_{5100}$ relation, suggesting that using the \CIV\ line
luminosity will not degrade the mass estimates much than using $L_{\rm
1350}$. The best-fit slopes of these correlations are slightly different from
unity, indicating a possible mild luminosity dependence of the ratios of
these luminosities to $L_{\rm 5100}$ over this luminosity range.

We note that these luminosity correlations are not predominately caused by
the common distance of each object. In fact, the dynamical range resulting
from luminosity distances is only 0.4 dex given the limited redshift range of
our objects, while the entire luminosity span is 1.5 dex. These luminosity
correlations justify the usage of alternative luminosity indicators in
various virial mass estimators. The scatter in the correlation between
alternative luminosity indicator and $L_{5100}$ will be one source of the
scatter in virial mass estimates when compared with those based on \hbeta\
width and $L_{5100}$.

\begin{deluxetable*}{lccc}
\tablecaption{Spectral Measurements \label{table:measure}} \tablehead{ &
Format & Units & Description} \startdata
objname & A10 & -- & Object Name \\
$\log L_{1350}$ & F6.3 & ${\rm erg\,s^{-1}}$ & continuum luminosity at restframe 1350\,\AA \\
Err $\log L_{1350}$ & F6.3 & ${\rm erg\,s^{-1}}$ & measurement error in $\log L_{1350}$ \\
$\log L_{3000}$ & F6.3 & ${\rm erg\,s^{-1}}$ & continuum luminosity at restframe 3000\,\AA \\
Err $\log L_{3000}$ & F6.3 & ${\rm erg\,s^{-1}}$ & measurement error in $\log L_{3000}$ \\
$\log L_{5100}$ & F6.3 & ${\rm erg\,s^{-1}}$ & continuum luminosity at restframe 5100\,\AA \\
Err $\log L_{5100}$ & F6.3 & ${\rm erg\,s^{-1}}$ & measurement error in $\log L_{5100}$ \\
$\log L_{\rm CIV}$ & F6.3 & ${\rm erg\,s^{-1}}$ & luminosity of the broad \CIV\ line \\
Err $\log L_{\rm CIV}$ & F6.3 & ${\rm erg\,s^{-1}}$ & measurement error in $\log L_{\rm CIV}$ \\
$\log L_{\rm CIII]}$ & F6.3 & ${\rm erg\,s^{-1}}$ & luminosity of the broad \CIII\ line \\
Err $\log L_{\rm CIII]}$ & F6.3 & ${\rm erg\,s^{-1}}$ & measurement error in $\log L_{\rm CIII]}$ \\
$\log L_{\rm MgII}$ & F6.3 & ${\rm erg\,s^{-1}}$ & luminosity of the broad \MgII\ line \\
Err $\log L_{\rm MgII}$ & F6.3 & ${\rm erg\,s^{-1}}$ & measurement error in $\log L_{\rm MgII}$ \\
$\log L_{\rm H\beta}$ & F6.3 & ${\rm erg\,s^{-1}}$ & luminosity of the broad \hbeta\ line \\
Err $\log L_{\rm H\beta}$ & F6.3 & ${\rm erg\,s^{-1}}$ & measurement error in $\log L_{\rm H\beta}$ \\
$\log L_{\rm H\alpha}$ & F6.3 & ${\rm erg\,s^{-1}}$ & luminosity of the broad \halpha\ line \\
Err $\log L_{\rm H\alpha}$ & F6.3 & ${\rm erg\,s^{-1}}$ & measurement error in $\log L_{\rm H\alpha}$ \\
${\rm FWHM_{CIV}}$ & I5 & ${\rm km\,s^{-1}}$ & FWHM of the broad \CIV\ line \\
Err ${\rm FWHM_{CIV}}$ & I5 & ${\rm km\,s^{-1}}$ & measurement error in ${\rm FWHM_{CIV}}$ \\
${\rm FWHM_{CIII]}}$ & I5 & ${\rm km\,s^{-1}}$ & FWHM of the broad \CIII\ line \\
Err ${\rm FWHM_{CIII]}}$ & I5 & ${\rm km\,s^{-1}}$ & measurement error in ${\rm FWHM_{CIII]}}$ \\
${\rm FWHM_{MgII}}$ & I5 & ${\rm km\,s^{-1}}$ & FWHM of the broad \MgII\ line \\
Err ${\rm FWHM_{MgII}}$ & I5 & ${\rm km\,s^{-1}}$ & measurement error in ${\rm FWHM_{MgII}}$ \\
${\rm FWHM_{H\beta}}$ & I5 & ${\rm km\,s^{-1}}$ & FWHM of the broad \hbeta\ line \\
Err ${\rm FWHM_{H\beta}}$ & I5 & ${\rm km\,s^{-1}}$ & measurement error in ${\rm FWHM_{H\beta}}$ \\
${\rm FWHM_{H\alpha}}$ & I5 & ${\rm km\,s^{-1}}$ & FWHM of the broad \halpha\ line \\
Err ${\rm FWHM_{H\alpha}}$ & I5 & ${\rm km\,s^{-1}}$ & measurement error in ${\rm FWHM_{H\alpha}}$ \\
$V_{\rm CIV-H\beta}$ & I5 & ${\rm km\,s^{-1}}$ & blueshift of the broad \CIV\ centroid w.r.t. the broad \hbeta\ centroid \\
Err $V_{\rm CIV-H\beta}$ & I5 & ${\rm km\,s^{-1}}$ & measurement error in $V_{\rm CIV-H\beta}$  \\
$V_{\rm CIV-AlIII}$ & I5 & ${\rm km\,s^{-1}}$ & blueshift of the broad \CIV\ centroid w.r.t. the broad \AlIII\ centroid \\
Err $V_{\rm CIV-AlIII}$ & I5 & ${\rm km\,s^{-1}}$ & measurement error in $V_{\rm CIV-AlIII}$  \\
$V_{\rm CIII]-H\beta}$ & I5 & ${\rm km\,s^{-1}}$ & blueshift of the broad \CIII\ centroid w.r.t. the broad \hbeta\ centroid \\
Err $V_{\rm CIII]-H\beta}$ & I5 & ${\rm km\,s^{-1}}$ & measurement error in $V_{\rm CIII]-H\beta}$  \\
$V_{\rm MgII-[OIII]}$ & I5 & ${\rm km\,s^{-1}}$ & blueshift of the broad \MgII\ centroid w.r.t. the narrow \OIII\ centroid \\
Err $V_{\rm MgII-[OIII]}$ & I5 & ${\rm km\,s^{-1}}$ & measurement error in $V_{\rm MgII-[OIII]}$  \\
$V_{\rm H\beta-[OIII]}$ & I5 & ${\rm km\,s^{-1}}$ & blueshift of the broad \hbeta\ centroid w.r.t. the narrow \OIII\ centroid \\
Err $V_{\rm H\beta-[OIII]}$ & I5 & ${\rm km\,s^{-1}}$ & measurement error in $V_{\rm H\beta-[OIII]}$  \\
$V_{\rm H\alpha-[OIII]}$ & I5 & ${\rm km\,s^{-1}}$ & blueshift of the broad \halpha\ centroid w.r.t. the narrow \OIII\ centroid \\
Err $V_{\rm H\alpha-[OIII]}$ & I5 & ${\rm km\,s^{-1}}$ & measurement error in $V_{\rm H\alpha-[OIII]}$  \\
CIV AS & F4.2 & -- & Asymmetry parameter of the broad \CIV\ line
\enddata
\tablecomments{Format of the tabulated spectral measurements. The full table
is available in the online version. }
\end{deluxetable*}

\begin{deluxetable}{lccc}
\tablecaption{BCES Bisector Slopes for Luminosity Correlations
\label{table:bces}} \tablehead{ vs $L_{5100}$ & $\alpha$ & $\sigma_{\alpha}$
& Scatter} \startdata
$L_{1350}$ & 1.044 & 0.099 & 0.13 \\
$L_{3000}$ & 0.979 & 0.036 & 0.05 \\
$L_{\rm H\alpha,broad}$ & 1.010 & 0.042 & 0.07 \\
$L_{\rm H\beta,broad}$ & 1.251 & 0.067 & 0.11 \\
$L_{\rm MgII,broad}$ & 0.861 & 0.070 & 0.11 \\
$L_{\rm CIII],broad}$ & 0.924 & 0.099 & 0.16 \\
$L_{\rm CIV,broad}$ & 0.950 & 0.085 & 0.14 \\
\enddata
\tablecomments{$\alpha$ and $\sigma_{\alpha}$ are the slope and uncertainty
(1$\sigma$) in slope from the bisector linear regression fit of each
luminosity against $L_{5100}$. The last column lists the scatter
perpendicular to the best-fit linear relation (dominated by intrinsic scatter
rather than measurement errors).}
\end{deluxetable}

\begin{deluxetable}{lcc}
\tablecaption{Spearman Test Results \label{table:spearman1}} \tablehead{ &
$r_s$ & $P_{\rm ran}$ } \startdata
with FWHM$_{\rm H\beta}$         & & \\
\\
FWHM$_{\rm CIV}$        & 0.11 & 0.39 \\
FWHM$_{\rm CIII]}$      & 0.14 & 0.29 \\
FWHM$_{\rm MgII}$       & 0.64 & $2.8\times 10^{-8}$ \\
FWHM$_{\rm H\alpha}$    & 0.78 & $1.8\times 10^{-13}$ \\
FWHM$_{\rm CIV,corr}$   & 0.49 & $6.7\times 10^{-5}$  \\
FWHM$_{\rm CIII],corr}$ & 0.36 & $5.2\times 10^{-3}$ \\
\hline\\
with $\frac{\rm FWHM_{\rm CIV}}{\rm FWHM_{H\beta}}$ & & \\
\\
$\log L_{1350}$         & 0.02  & 0.89 \\
$\log L_{5100}$         & -0.03 & 0.83 \\
$\log (L_{1350}/L_{5100})$ & 0.21 & 0.11 \\
EW \CIV\                & -0.07  & 0.59 \\
$V_{\rm CIV-H\beta}$    & 0.59  & $8.9\times 10^{-7}$ \\
$V_{\rm CIV-AlIII}$     & 0.36 & $5.2\times 10^{-3}$  \\
\CIV\ AS                & -0.28 & 0.028
\enddata
\tablecomments{$r_s$ is the Spearman rank-order coefficient and $P_{\rm ran}$
is the probability of being drawn from random distributions. FWHM$_{\rm
CIV,corr}$ and FWHM$_{\rm CIII],corr}$ are the corrected FWHMs using the
linear regression fits shown in Fig.\ \ref{fig:ciii_civ_fwhm_corr}.}
\end{deluxetable}


\subsection{Line Width Correlations}\label{subsec:corr_line}

\begin{figure}
 \centering
 \includegraphics[width=0.48\textwidth]{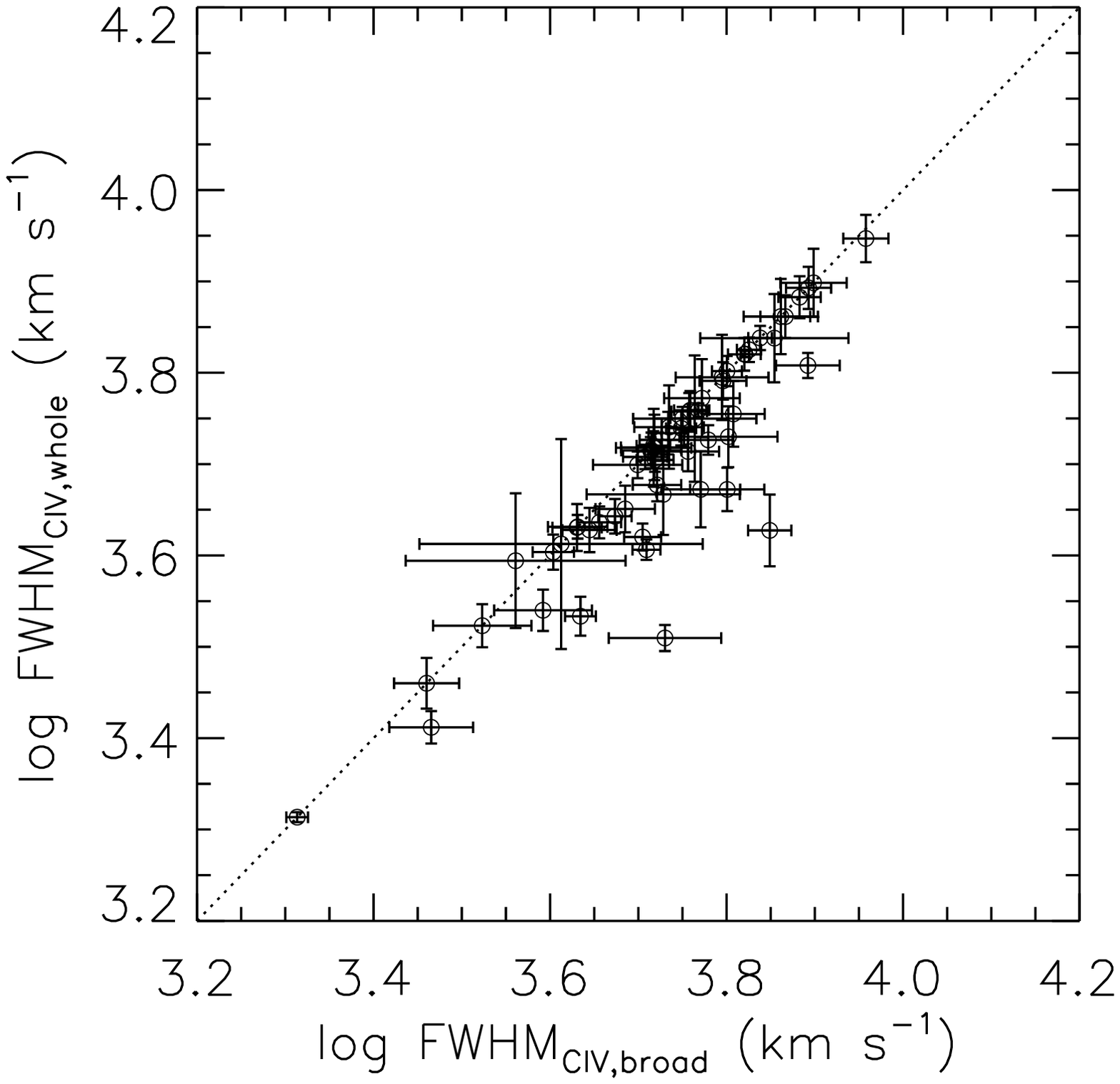}
 \includegraphics[width=0.48\textwidth]{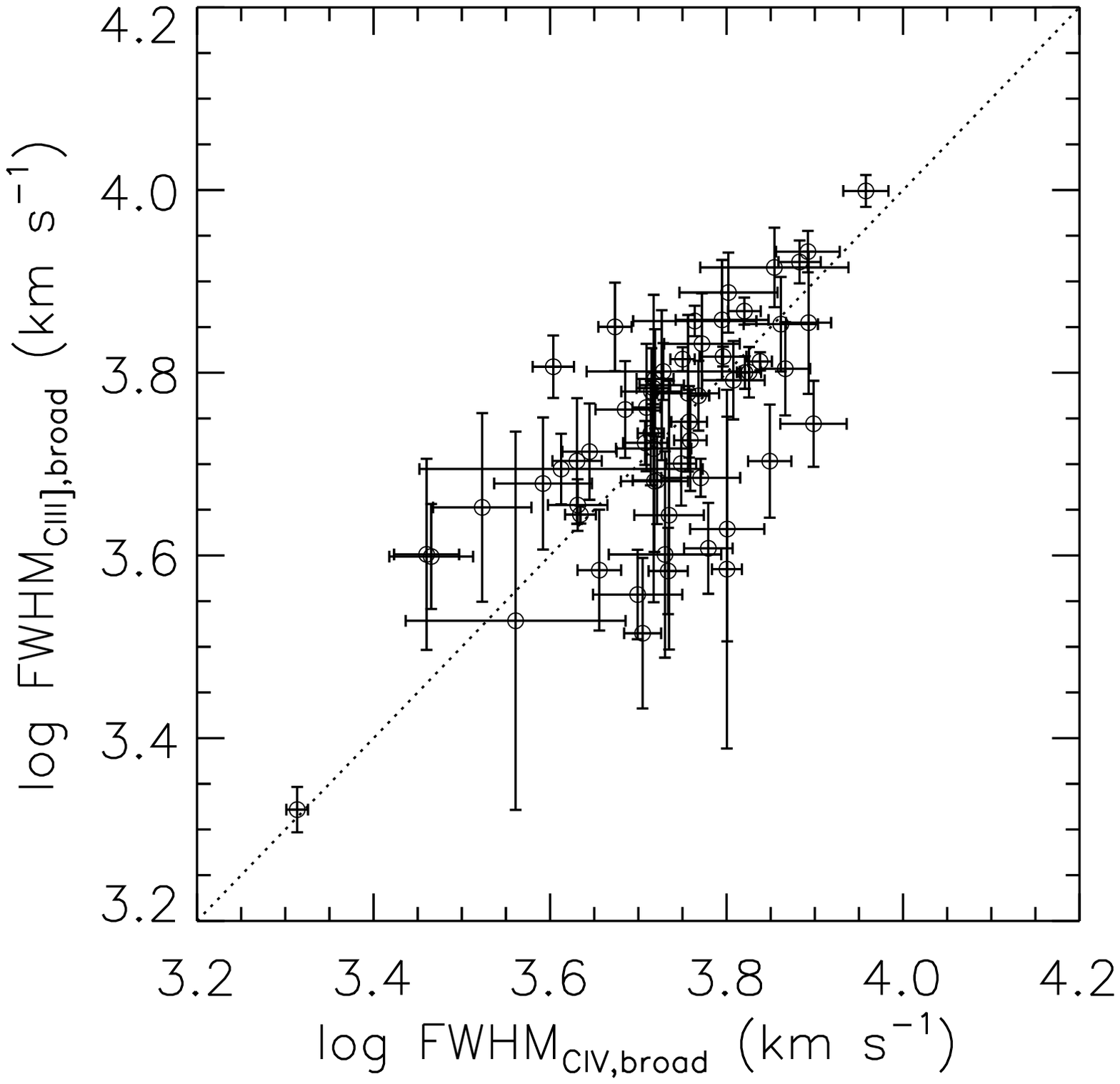}
    \caption{{\em Upper:} Comparison between the two methods of measuring the
    \CIV\ FWHM, i.e., with and without subtracting a narrow line component. The two
    methods yield similar results for the majority of our objects, indicating that
    the narrow line contribution is generally negligible for \CIV\ for luminous quasars
    with $L_{5100}>10^{45.4}\,{\rm erg\,s^{-1}}$. {\em Bottom:} Comparison between
    the \CIV\ and \CIII\ FWHMs. The uncertainties associated with the \CIII\ FWHM
    measurements are typically large due to ambiguities in decomposing the \CIII\
    complex, but a general correlation is seen between the two FWHMs. }\label{fig:civ_fwhm}
\end{figure}

\begin{figure}
 \centering
 \includegraphics[width=0.48\textwidth]{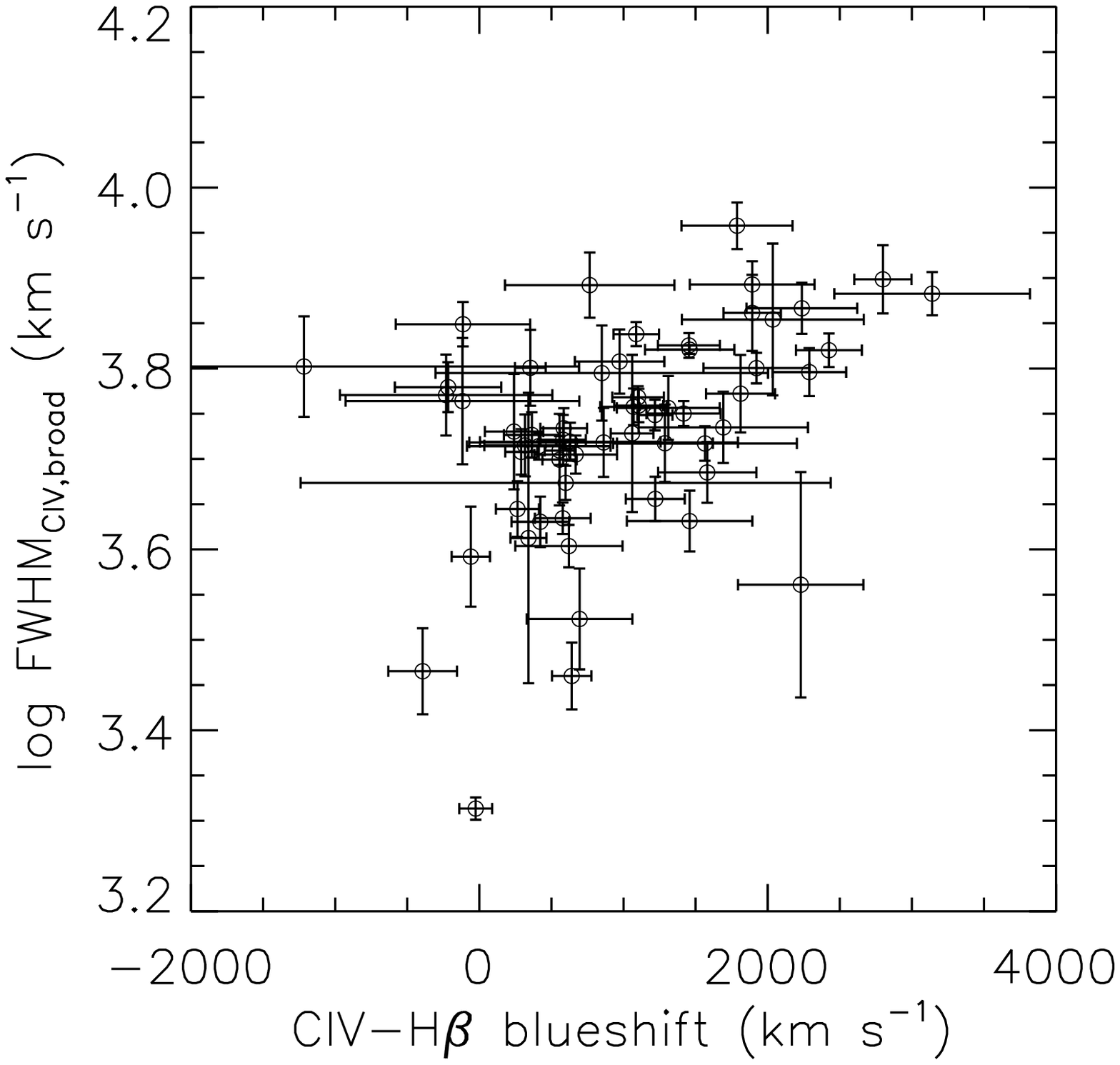}
 \includegraphics[width=0.48\textwidth]{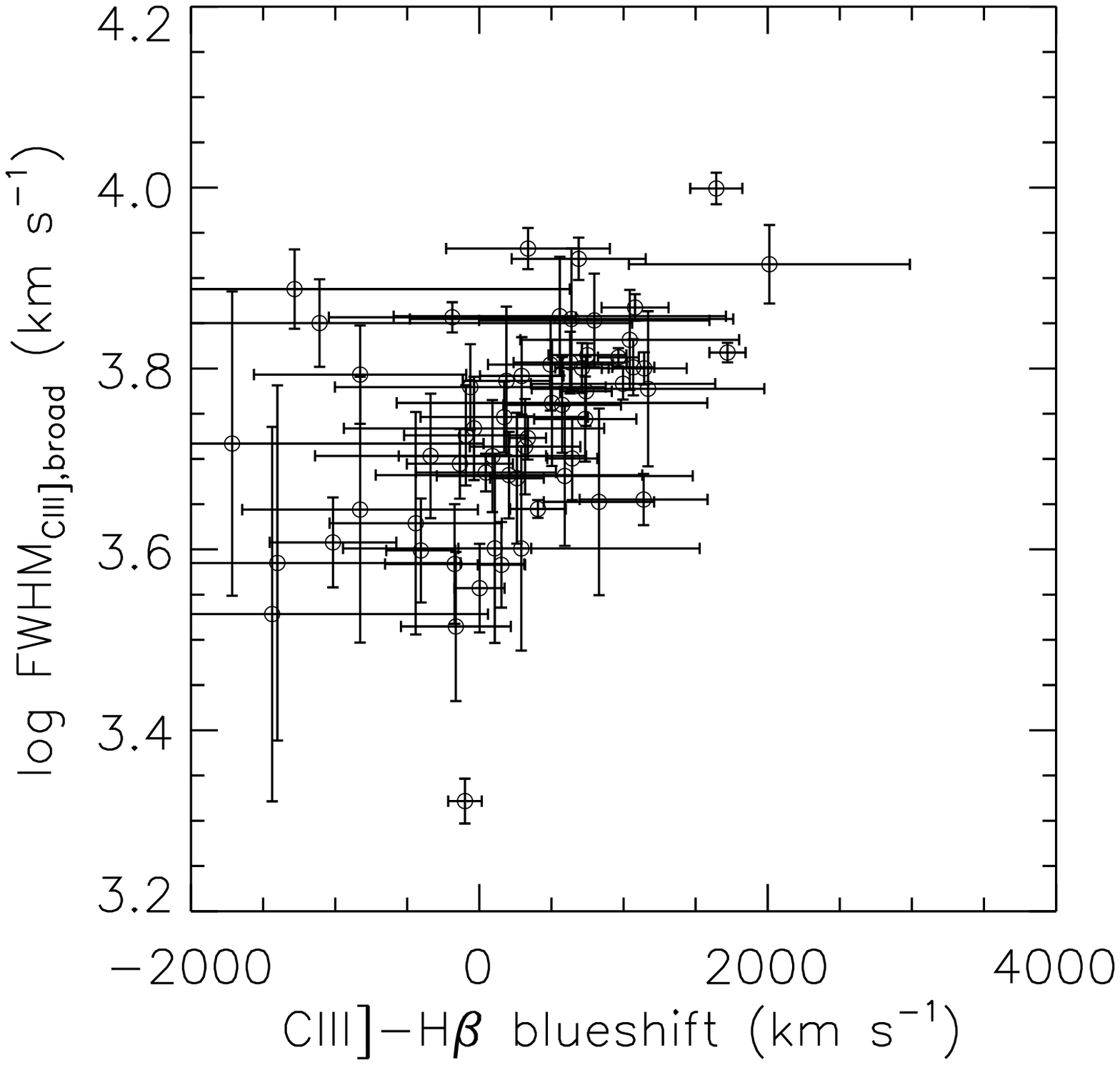}
    \caption{FWHM as a function of the blueshift relative to the broad \hbeta\
    centroid for \CIV\ (upper) and \CIII\ (bottom). Both FWHMs seem to increase
    with the blueshift relative to \hbeta. Spearman correlation test
    results are $r_s=0.44$ ($P_{\rm ran}=5.1\times10^{-4}$) for \CIV\ and
    $r_s=0.50$ ($P_{\rm ran}=4.7\times10^{-5}$) for \CIII.}\label{fig:civ_voff}
\end{figure}

While it is still debated whether or not a narrow line component for \CIV\
needs to be subtracted for high-redshift broad-line quasars \citep[see
discussions in, e.g., ][]{Bachev_etal_2004,Sulentic_etal_2007}, it is clear
that narrow \CIV\ emission does exist, as seen in some type 2 quasars
\citep[e.g.,][]{Stern_etal_2002}. The \OIII\ coverage in our near-IR spectra
makes it possible to constrain the strength of the narrow \CIV\ emission by
fixing its line width and velocity offset to those of the narrow \OIII\
lines. We have measured the width of \CIV\ with and without the subtraction
of a possible narrow line component. In Fig.\ \ref{fig:civ_fwhm} we compare
the resulting \CIV\ FWHM with the two methods. The two objects with large
error bars (J1009$+$0230 and J1542$+$1112) have associated absorption, which
causes some ambiguities in decomposing the \CIV\ line in our Monte Carlo mock
spectra and therefore leads to large uncertainties (see
\S\ref{sec:spec_measure}). We found that the narrow line contribution to
\CIV\ is generally weak for objects in our sample, and only in 2 objects
(J1119$+$2332 and J1710$+$6023) the narrow \CIV\ component is strong enough
to make a large difference in the line width measurement. From now on we use
the \CIV\ line width with narrow line subtraction. We note, however, that our
quasars are luminous, and the relative strength of the narrow \CIV\ emission
may be larger for lower luminosity objects \citep[see,
e.g.,][]{Bachev_etal_2004,Sulentic_etal_2007}.

In the bottom panel of Fig.\ \ref{fig:civ_fwhm} we compare the \CIV\ FWHM and
\CIII\ FWHM. The measurement errors are typically larger for \CIII\ due to
the ambiguity of decomposing the \CIII\ complex, but a correlation is still
seen between the FWHMs of \CIV\ and \CIII. This is intriguing because \CIII\
does not show as large a blueshift relative to the low-ionization lines as
does \CIV\ \citep[e.g.,][Shen et~al., in preparation]{Richards_etal_2011}
when the contributions from \SiIII\ and \AlIII\ are removed. In Fig.\
\ref{fig:civ_voff} we plot the FWHM against the velocity offset relative to
the broad \hbeta\ line, for \CIV\ and \CIII\ respectively. In both cases a
significant positive correlation is detected, i.e., FWHM increases with
blueshift. Such a trend is already known when comparing \CIV\ and \MgII\
\citep[e.g.,][]{Shen_etal_2008b,Shen_etal_2011}, and now it is confirmed when
comparing \CIV\ directly with \hbeta. However, similar trends were not found
for the FWHM of \halpha, \hbeta\ or \MgII\ against their velocity shift
relative to \OIII. We also tested if there is any correlation between FWHM
and other properties (continuum luminosity, color, line asymmetry) for all
five lines and found none of them is significant.

\begin{figure*}
 \centering
 \includegraphics[width=0.98\textwidth]{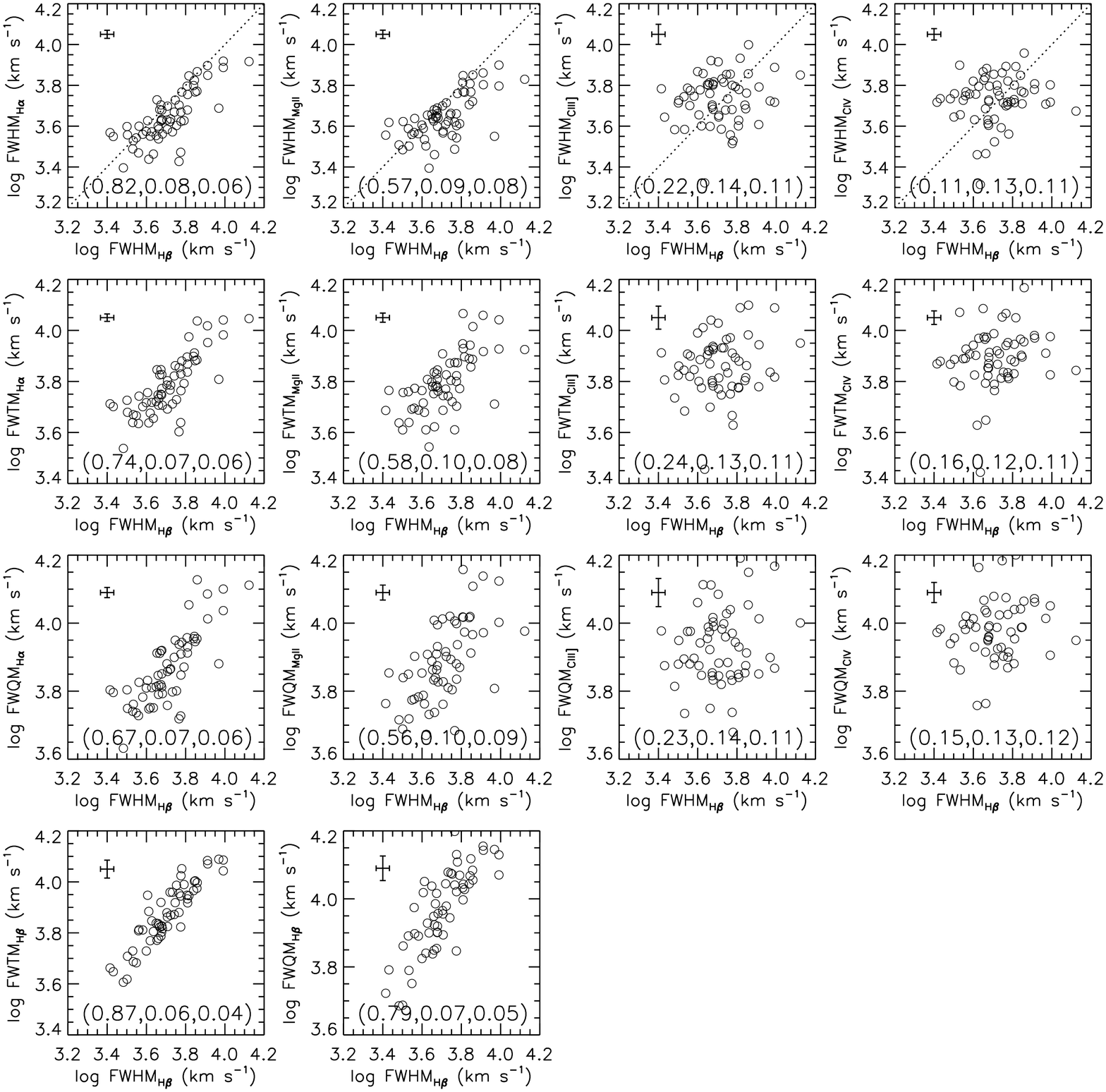}
    \caption{Comparisons between different line widths and \hbeta\ FWHM. Typical measurement uncertainties are
    indicated in each panel. The first row
    compares the FWHMs of \halpha, \MgII, \CIII\ and \CIV\ with \hbeta\ FWHM, where the
    dotted lines show the unity relation. Only the FWHMs of \halpha\ and
    \MgII\ show significant correlation with the FWHM of \hbeta. The second
    and third rows show similar comparisons, but with FWTMs and FWQMs for
    \halpha, \MgII\, \CIII\ and \CIV. The conclusion is the same. The bottom two panels show the
    correlations between FWTM/FWQM and FWHM for \hbeta. In each panel we show the best-fit
    linear regression results using the Bayesian method in \citet[][predicting Y at X]{Kelly_2007}:
    the best-fit slope, the uncertainty of the slope and the intrinsic random scatter about the regression.}\label{fig:line_width_all}
\end{figure*}

In Fig.\ \ref{fig:line_width_all} we plot different line widths against the
broad \hbeta\ FWHM. We have suppressed measurement errors in these plots for
clarity. In addition to the traditional FWHM, we also measure the
full-width-at-third-maximum (FWTM) and full-width-at-quarter-maximum (FWQM)
as alternative line width indicators. Consistent with earlier studies, we see
strong correlations among the widths of \halpha, \hbeta\ and \MgII\
\citep[e.g.,][]{Greene_Ho_2005,Salviander_etal_2007,Shen_etal_2008b,Wang_etal_2009b}.
On the other hand, both \CIII\ and \CIV\ line widths show poor correlations
with \hbeta\ line width. Table \ref{table:spearman1} lists the Spearman
rank-order coefficients of these correlations. Since we found using FWTM and
FWQM does not improve the correlations, we will focus on FWHM from now on.

\begin{figure*}
 \centering
 \includegraphics[width=0.46\textwidth]{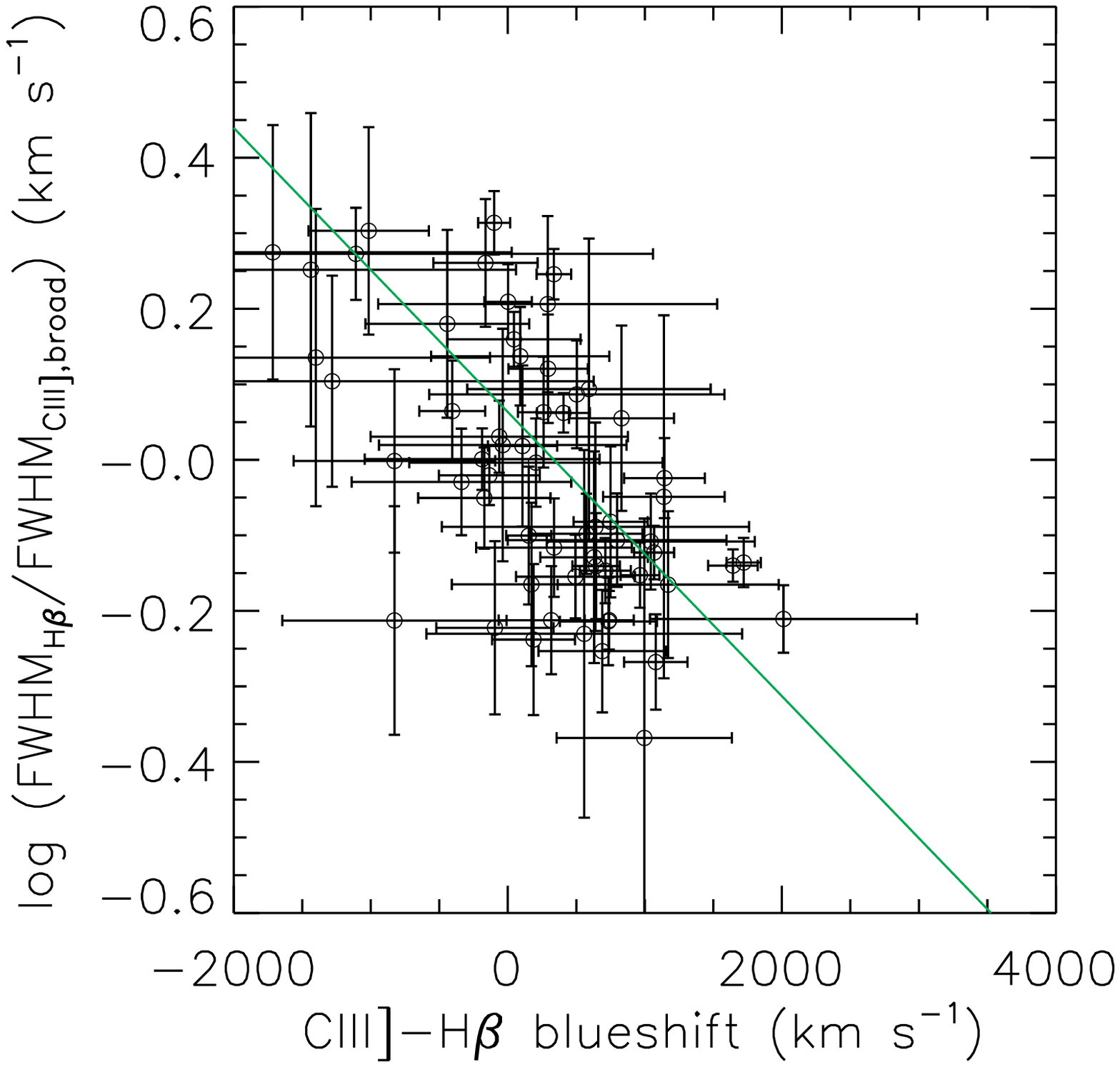}
 \includegraphics[width=0.46\textwidth]{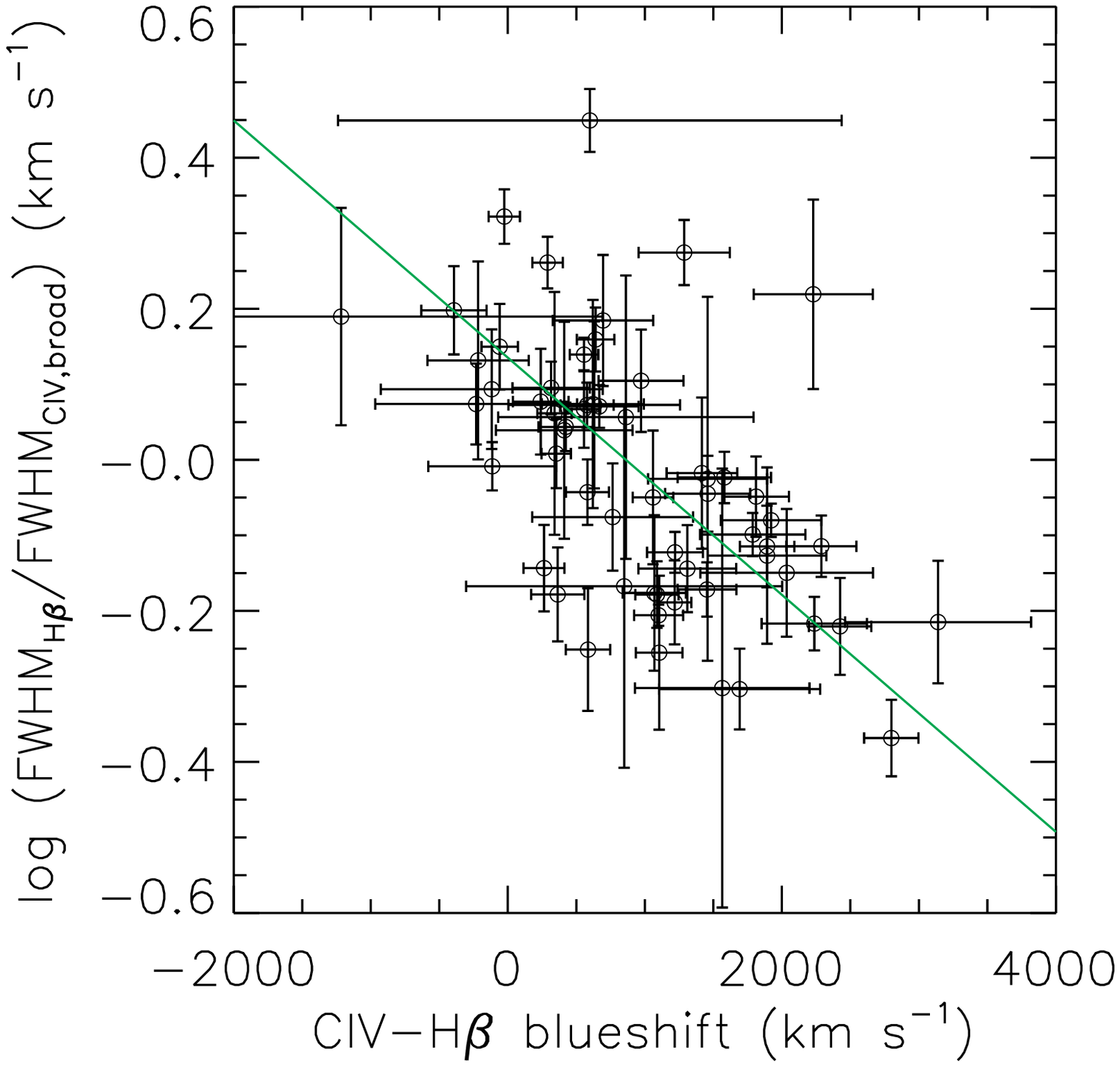}
 \includegraphics[width=0.46\textwidth]{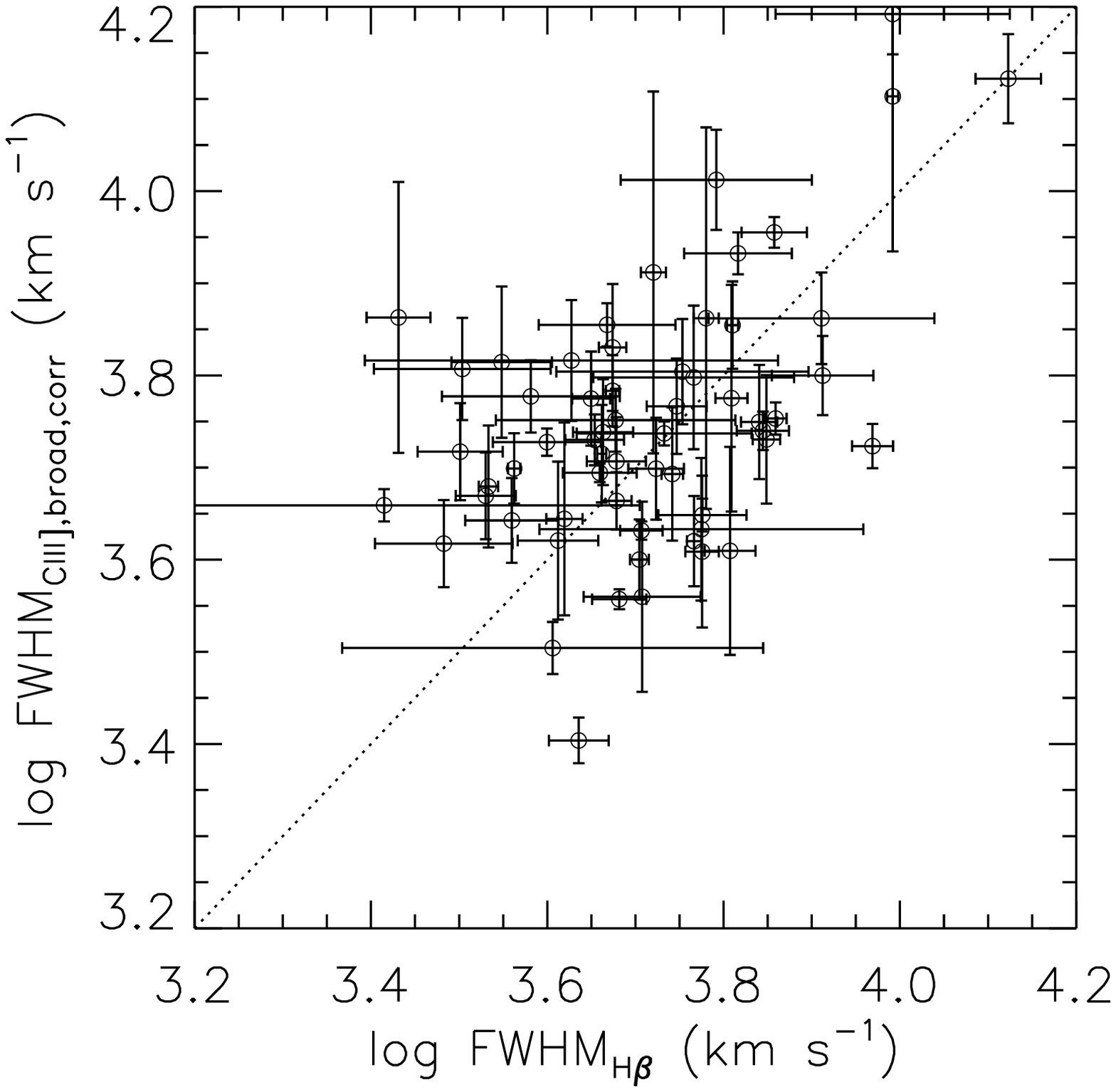}
 \includegraphics[width=0.46\textwidth]{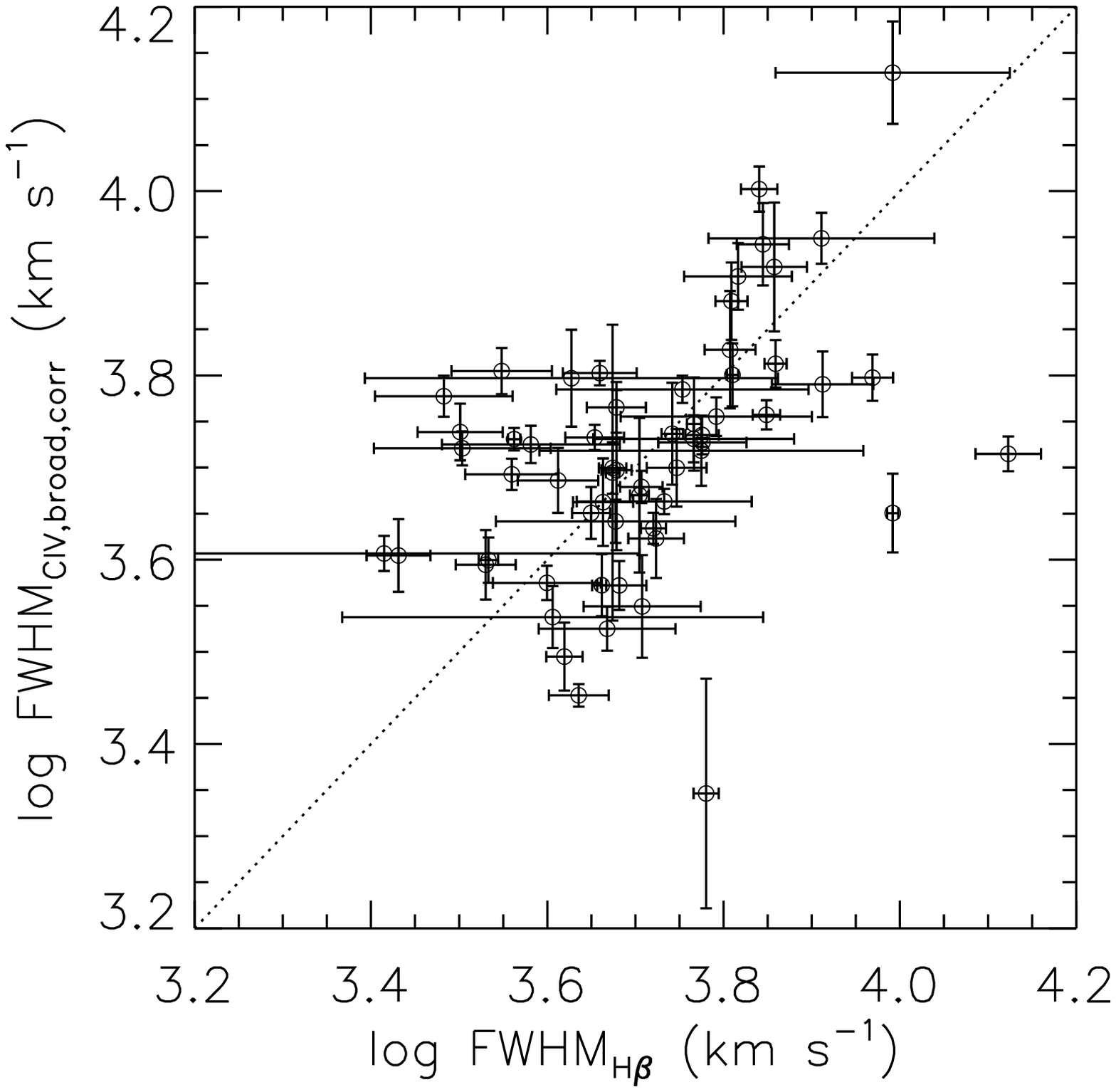}
    \caption{{\em Top:} FWHM ratio as a function of blueshift, for \CIII\
    (left) and \CIV\ (right), compared with \hbeta. A significant correlation
    is detected for both \CIII\ and \CIV. The green lines are the best linear
    regression fits using the Bayesian method in \citet{Kelly_2007}. {\em Bottom:}
    ``Corrected'' FWHMs for \CIII\ (left) and \CIV\ (right) using the best
    fits shown in the upper panels, compared with \hbeta\ FWHM. A better
    one-to-one correlation is now seen between the \CIII/\CIV\ FWHM and
    \hbeta\ FWHM, although significant scatter still remains.
    }\label{fig:ciii_civ_fwhm_corr}
\end{figure*}

These results suggest that \CIV\ and \CIII\ have different kinematics from
\MgII\ and the Balmer lines, and possibly originate from a different region
than the low-ionization lines. However, since some dispersion in the \CIV\
and \CIII\ FWHM is driven by the blueshift (e.g., Fig.\ \ref{fig:civ_voff}),
accounting for this dependence may reduce the difference in FWHM between
\CIV/\CIII\ and the low-ionization lines. To test this, we plot the
difference in FWHM $\Delta_{\rm FWHM}=\log({\rm
FWHM_{H\beta}/FWHM_{CIII],CIV}})$, as a function of the blueshift with
respect to \hbeta for \CIII\ and \CIV, in Fig.\ \ref{fig:ciii_civ_fwhm_corr}
(top panels). The green lines are the linear regression results using the
Bayesian method of \citet[][]{Kelly_2007}. We then use these best-fit linear
relations to correct for the observed \CIII\ and \CIV\ FWHMs:
\begin{equation}
\log\,{\rm FWHM_{CIII]/CIV,corr}}=\log\,{\rm FWHM_{CIII]/CIV}} + \alpha + \beta\Delta V\ ,
\end{equation}
where $\Delta V=(v_{\rm off, CIII]/CIV}-v_{\rm off,H\beta})$ is the blueshift
relative to \hbeta, and $\alpha$ and $\beta$ are the best-fit coefficients of
the linear regression shown in green lines in the top panels of Fig.\
\ref{fig:ciii_civ_fwhm_corr}; $[\alpha,\beta]=[0.062,-1.87\times10^{-4}]$ for
\CIII, and $[\alpha,\beta]=[0.136,-1.57\times10^{-4}]$ for \CIV. The
``corrected'' \CIII\ and \CIV\ FWHMs are plotted against the broad \hbeta\
FWHMs in the bottom panels of Fig.\ \ref{fig:ciii_civ_fwhm_corr}. This time
significant correlations are detected for both \CIII\ and \CIV, and \CIV\ has
the most significant improvement (see Table \ref{table:spearman1} for
Spearman rank-order coefficients). Nevertheless, there is still substantial
scatter ($\sim 0.15$ dex for \CIII\ and $\sim 0.12$ dex for \CIV) among these
correlations. This ``irreducible'' scatter probably again reflects the
different origins of \CIII\ and \CIV\ from the low-ionization lines, which
makes it difficult to bring their line widths into good agreement. We will
return to this point in \S\ref{subsec:comp}.

Although the blueshift relative to \hbeta\ seems a viable proxy to correct
the \CIII\ and \CIV\ FWHM to better agree with the \hbeta\ FWHM, it is of
little practical value. One would like a proxy that can be determined from
regions around the \CIII\ or \CIV\ line alone. We have tried to correlate
$\Delta_{\rm FWHM}$ with $\log L_{1350}$, ${\rm EW_{CIV}}$, asymmetry
parameters of \CIV\ (defined as $\displaystyle{\rm AS}\equiv
\ln\left(\frac{\lambda_{\rm red}}{\lambda_0}\right)/\ln
\left(\frac{\lambda_0}{\lambda_{\rm blue}}\right)$, where $\lambda_0$ is the
peak flux wavelength, and $\lambda_{\rm red}$ and $\lambda_{\rm blue}$ are
the wavelengths at half peak flux from the model fits), and blueshift
relative to \AlIII. We found that $\Delta_{\rm FWHM}$ is best correlated with
asymmetry parameters of \CIV\ and the blueshift relative to \AlIII\ at the
$P_{\rm ran}<10^{-2}$ level, although still worse than the ones against the
blueshift relative to \hbeta\ (where $P_{\rm ran}<10^{-6}$). Using these weak
correlations to correct for \CIV\ FWHM does not seem to reduce the scatter
between the \CIV\ and \hbeta\ FWHM much.

\begin{deluxetable}{lcccccc}
\tablecaption{Virial Mass Calibrations \label{table:linreg}} \tablehead{ Type
& $a$ & $b$ & $c$ & offset & $\sigma$ & Ref\\
& & & & (dex) & (dex) & } \startdata
Fiducial mass & & & & & & \\
${\rm FWHM_{H\beta}},L_{5100}$         & 0.91 & 0.5 & 2 & -- & -- & VP06 \\
\hline
Previous calibrations & & & & & \\\\
${\rm FWHM_{H\alpha}},L_{5100}$        & 0.774 & 0.520 & 2.06 & 0.01 & 0.14 & A11 \\
${\rm FWHM_{H\alpha}},L_{\rm H\alpha}$ & 1.239 & 0.430 & 2.10 & $-0.08$ & 0.15 & S11 \\
${\rm FWHM_{H\beta}},L_{5100}$         & 0.895 & 0.520 & 2.00 &  0.03 & 0.003 & A11 \\
${\rm FWHM_{H\beta}},L_{\rm H\beta}$   & 1.676 & 0.560 & 2.00 & $-0.04$ & 0.07 & GH05 \\
${\rm FWHM_{MgII}},L_{3000}$           & 0.860 & 0.500 & 2.00 & 0.02 & 0.16  & VO09$^*$ \\
${\rm FWHM_{MgII}},L_{3000}$           & 0.740 & 0.620 & 2.00 & 0.17 & 0.17  & S11$^*$ \\
${\rm FWHM_{MgII}},L_{\rm MgII}$       & 1.933 & 0.630 & 2.00 & 0.22 & 0.16  & S11$^*$ \\
${\rm FWHM_{CIV}},L_{1350}$            & 0.660 & 0.530 & 2.00 & 0.10  & 0.40 & VP06 \\
${\rm FWHM_{CIV}},L_{\rm CIV}$         & 1.525 & 0.457 & 2.00 & 0.09 & 0.38 & S11 \\
\hline
This work & & & & & & \\
($L_{5100}>10^{45.4}\,{\rm erg\,s^{-1}}$) & & & & & & \\\\
${\rm FWHM_{H\alpha}},L_{5100}$        & 1.390 & 0.555 & 1.873 & 0.01 & 0.12
& -- \\ 
${\rm FWHM_{H\alpha}},L_{\rm H\alpha}$ & 2.216 & 0.564 & 1.821 & 0.008 & 0.15 & -- \\ 
${\rm FWHM_{H\beta}},L_{\rm H\beta}$   & 1.963 & 0.401 & 1.959 & 0.02 & 0.04 & -- \\ 
${\rm FWHM_{MgII}},L_{3000}$           & 1.816 & 0.584 & 1.712 & 0.02 & 0.16 & -- \\ 
${\rm FWHM_{MgII}},L_{\rm MgII}$       & 3.979 & 0.698 & 1.382 & 0.03 & 0.15 & -- \\ 
${\rm FWHM_{CIV}},L_{1350}$            & 7.295 & 0.471 & 0.242 & 0.03 & 0.28 & -- \\ 
${\rm FWHM_{CIV}},L_{\rm CIV}$         & 7.535 & 0.639 & 0.319 & 0.002 & 0.26 & -- \\\\ 
$+z<0.9$ SDSS quasars & & & & & & \\
($L_{5100}>10^{45}\,{\rm erg\,s^{-1}}$) & & & & & & \\\\
${\rm FWHM_{MgII}},L_{3000}$           & 0.963 & 0.468 & 2.010 & 0.02 & 0.16 & -- \\
\enddata
\tablecomments{Virial BH mass calibrations of Eq.\ (\ref{eqn:calib}) based on
different line width and luminosity combinations for the 60 objects in our
sample, calibrated against the VP06 FWHM-based \hbeta\ virial masses.
References of previous calibrations: GH05 \citep{Greene_Ho_2005}, VP06
\citep{Vestergaard_Peterson_2006}, VO09 \citep{Vestergaard_Osmer_2009}, S11
\citep{Shen_etal_2011}, A11 \citep{Assef_etal_2011}. For each calibration we
measure the average offset and scatter $\sigma$ relative to the fiducial mass
$\log M_{\rm fid}$ by fitting a Gaussian to the mass residual $\Delta = \log
M-\log M_{\rm fid}$. The scatter in the mass residual is dominated by
intrinsic scatter than by measurement errors. $^*$In using the \MgII\
calibrations in \citet{Shen_etal_2011} and \citet{Vestergaard_Osmer_2009} we
have added $0.125$\,dex to the measured $\log L_{\rm 3000}$ and subtracted
$0.067$\, dex from the measured $\log L_{\rm MgII}$ to account for the
different fitting recipe used in this work. Note that the absolute
uncertainties of these calibrations are on the level of $>0.3$\ dex
\citep[e.g.,][]{Peterson_2011}.}
\end{deluxetable}

\subsection{Comparing Single-Epoch Virial Mass Estimators}

\begin{figure*}
 \centering
 \includegraphics[width=0.8\textwidth]{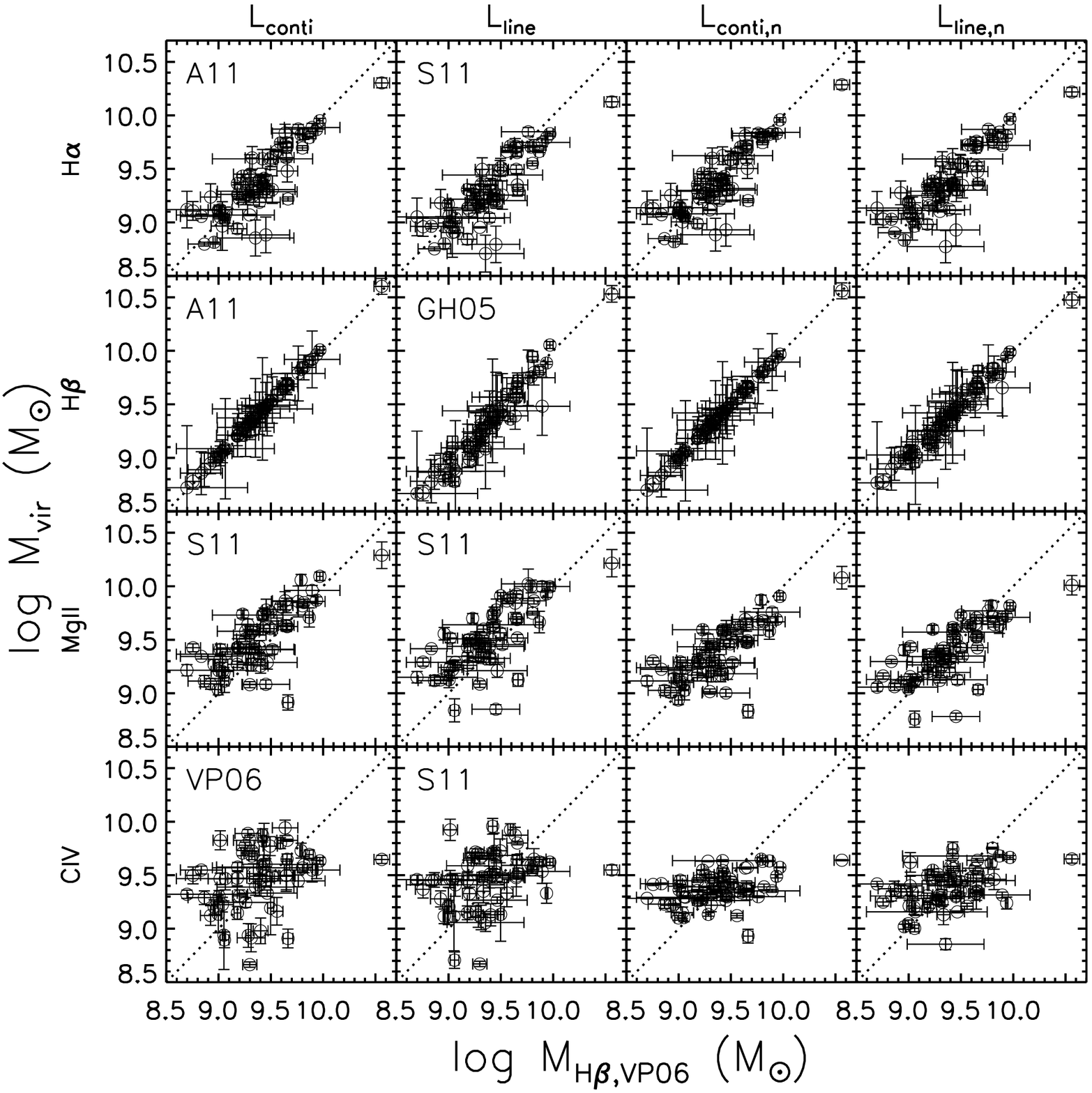}
    \caption{Comparisons between different virial mass estimators with the \hbeta\
    estimator in \citet{Vestergaard_Peterson_2006}. The left two columns show the
    comparisons for existing calibrations based on either continuum luminosity or
    line luminosity, from \citet[][A11]{Assef_etal_2011}, \citet[][S11]{Shen_etal_2011},
    \citet[][GH05]{Greene_Ho_2005}, and \citet[][VP06]{Vestergaard_Peterson_2006}. The right two columns show the results for our new calibrations
    based on our sample, where we allow the slope on FWHM to vary for
    alternative lines to minimize the residual in virial mass estimates when
    compared to \hbeta. The linear regression results are listed in Table
    \ref{table:linreg}.
    }\label{fig:mass_comp}
\end{figure*}

The investigations so far in the previous two sections treated luminosity and
line width independently. In principle, if there is covariance between line
width and luminosity when comparing the virial products based on different
lines, the resulting scatter in the residual virial products may be increased
or reduced. Since we did not observe any strong dependence of line width on
luminosity for any particular line, we expect such effects to be modest at
most.

As reasoned in the introduction, we adopt the \hbeta$+L_{5100}$ virial masses
as the ``truth'' values, and minimize the differences using alternative line
estimator with respect to \hbeta. There is more than one calibration based on
$L_{5100}$ and ${\rm FWHM_{H\beta}}$
\citep[e.g.,][]{Mclure_Dunlop_2004,Vestergaard_Peterson_2006,Assef_etal_2011},
and we use the \citet[][VP06]{Vestergaard_Peterson_2006} calibration as our
standard, which is compatible with our measurements of the \hbeta\ FWHM, and
provides similar estimates to those using the calibration in \citet[][see
below]{Assef_etal_2011}. The virial mass estimator based on a particular pair
of line width and luminosity is:
\begin{equation}\label{eqn:calib}
\log\left(\frac{M_{\rm BH,vir}}{M_\odot}\right)= a + b\log\left(\frac{L}{10^{44}\, {\rm erg\,s^{-1}}}\right) + c\log\left(\frac{\rm FWHM}{\rm
km\,s^{-1}}\right)\ ,
\end{equation}
where $L$ and FWHM are the continuum (or line) luminosity and width for the
specific line, and coefficients $a$, $b$ and $c$ are to be determined by
linear regression analysis.

In order to minimize the difference in virial masses compared to our fiducial
masses, we allow the slopes on both luminosity and FWHM to vary. We use the
multi-dimensional Bayesian linear regression method in \citet{Kelly_2007} to
perform regression, treating our standard masses as the dependent variable
$Y$, and $(\log L,\log{\rm FWHM})$ as the 2-dimensional independent variable
$X$. This approach takes into account measurement errors, and possible
covariance between luminosity and FWHM, thus is arguably better than
regressions on $L$ versus $L_{5100}$ and ${\rm FWHM}$ versus ${\rm
FWHM_{H\beta}}$ separately. The regression results are listed in Table
\ref{table:linreg}.

In Fig.\ \ref{fig:mass_comp} we show the comparisons between different virial
mass estimators and the VP06 \hbeta\ estimator. We show in the first two
columns the comparisons for several calibrations in earlier work, and in the
last two columns the comparisons for our new calibrations as summarized in
Table \ref{table:linreg}.

These earlier calibrations were calibrated using fainter samples than probed
here, but in general they provide mass estimates that agree with the fiducial
\hbeta-based virial masses (mean offset $\lesssim 0.1$\,dex). The exceptions
are the \MgII-based calibrations in \citet{Shen_etal_2011}, where the steeper
slope of the $R-L$ relation for \MgII\ than for \hbeta\ has led to
increasingly larger discrepancies towards high luminosities\footnote{We note
that the \MgII\ calibrations in \citet{Shen_etal_2011} were not based on
linear regression fits against \hbeta\ masses, and had a slope in the $R-L$
relation fixed to be the one in \citet{Mclure_Dunlop_2004}. Using a steeper
slope 0.62 in $R-L_{3000}$ relation, the \MgII\ virial masses in
\citet{Shen_etal_2011} have negligible systematic offset relative to both
\hbeta-based masses at $z<0.9$ and \CIV-based masses at $z>1.5$. }. For our
new calibrations, the slope on FWHM is close to (albeit slightly smaller
than) 2 for \halpha, \hbeta\ and \MgII, indicating that using FWHM in these
calibrations improves the agreement with our standard mass estimator
(VP06-\hbeta). However, for \CIV, our linear regression result has a slope on
FWHM that is much shallower. This is because \CIV\ FWHM is poorly correlated
with \hbeta\ FWHM for our sample, and the scatter between the two FWHMs
rather than between the two continuum luminosities is the dominant source of
the difference in their virial masses \citep[in contrast to ][see discussions
in \S\ref{subsec:comp}]{Assef_etal_2011}; therefore the regression prefers a
smaller dependence on \CIV\ FWHM to minimize the difference in the two mass
estimates. In other words, the individual \CIV\ FWHM adds little to improve
the agreement with our standard mass estimates, but instead degrades the
agreement. We found similar trends for \CIII\ (not shown) as for \CIV.

We test the dependence of the \CIV\ virial mass residual on continuum
luminosity, color and line shifts. The mass residual is correlated with $\log
(L_{1350}/L_{5000})$, but this is expected since both masses involve
luminosity. Correcting for this color dependence only marginally improves the
agreement between the two masses. On the other hand, the dependence on
\CIV-\hbeta\ blueshift is strong enough such that incorporating this
dependence can improve the agreement between \CIV\ masses and the standard
masses. These results again reflect the fact that the difference in FWHM is
the dominant source in the virial mass difference. But this correction based
on the \CIV-\hbeta\ blueshift is of little practical use since there is no
need to correct \CIV-based virial masses if we have \hbeta\ coverage. Using
the \CIV-\AlIII\ blueshift or \CIV\ asymmetry as a surrogate for the
\CIV-\hbeta\ blueshift only leads to marginal improvement of the \CIV-based
masses, and thus is not of much practical value either.

Our new calibrations for \MgII\ yield consistent mass estimates as those
estimated from \hbeta\ for luminous ($L_{5100}>10^{45.4}\,{\rm erg\,s^{-1}}$)
quasars. However, it would be useful to derive a \MgII\ calibration that is
also applicable to lower luminosities. For this purpose we select $\sim 900$
$z<0.89$ quasars from the compilation in \citet{Shen_etal_2011} with good
\hbeta\ and \MgII\ measurements ($\Delta M_{\rm vir}<0.1$ dex) and
$L_{5100}>10^{45}\,{\rm erg\,s^{-1}}$ (to reduce host contamination). We
combine these quasars with the $60$ high-luminosity quasars in our sample and
perform the two-dimensional linear regression on $(\log L_{3000},\log{\rm
FWHM_{MgII}})$ against the fiducial \hbeta-based masses. The best-fit
coefficients are listed in Table \ref{table:linreg}. This \MgII\ calibration
is close to the one presented in \citet{Vestergaard_Osmer_2009}, and works
reasonably well for the entire luminosity regime
$10^{45}<L_{5100}<10^{47}\,{\rm erg\,s^{-1}}$.

\section{Discussion}\label{sec:dis}

\subsection{Comparison with Earlier Studies}\label{subsec:comp}

There have been many studies comparing different virial mass estimators
\citep[e.g.,][]{Vestergaard_Peterson_2006,
McGill_etal_2008,Dietrich_Hamann_2004,Dietrich_etal_2009,Netzer_etal_2007,Shen_etal_2008b,Shen_etal_2011,Wang_etal_2009b}.
These comparison studies used samples that have different sizes, spectral
quality, luminosity and redshift ranges, and focused on different lines. The
current work is among the few studies that simultaneously investigate \CIV\
through \halpha\ in the same objects, and our sample is substantially larger
and more homogeneous than used in similar studies
\citep[e.g.,][]{Dietrich_Hamann_2004,Dietrich_etal_2009}.

Our results agree with earlier work that the line width of \MgII\ is well
correlated with that of \hbeta\
\citep[e.g.,][]{Salviander_etal_2007,McGill_etal_2008,Shen_etal_2008b}, but
now we have extended this conclusion to higher luminosity than can be probed
with earlier samples at lower redshift. On the other hand, we confirm the
poor correlation between \CIV\ FWHM and \hbeta\ FWHM reported earlier
\citep[e.g.,][]{Baskin_Laor_2005,Netzer_etal_2007}, which was also inferred
from the comparison between \CIV\ and \MgII\ using SDSS quasars
\citep[e.g.,][]{Shen_etal_2008b}.

\citet{Assef_etal_2011} used a sample of high-redshift quasars with optical
(covering \CIV) and near-IR (covering the Balmer lines) spectroscopy to show
that there is a correlation between the widths of \CIV\ and the Balmer lines.
They further concluded that the correlation persists, although it becomes
weaker, when other objects compiled from \citet{Vestergaard_Peterson_2006},
\citet{Netzer_etal_2007} and \citet{Dietrich_etal_2009} are included.

\begin{figure}
 \centering
 \includegraphics[width=0.48\textwidth]{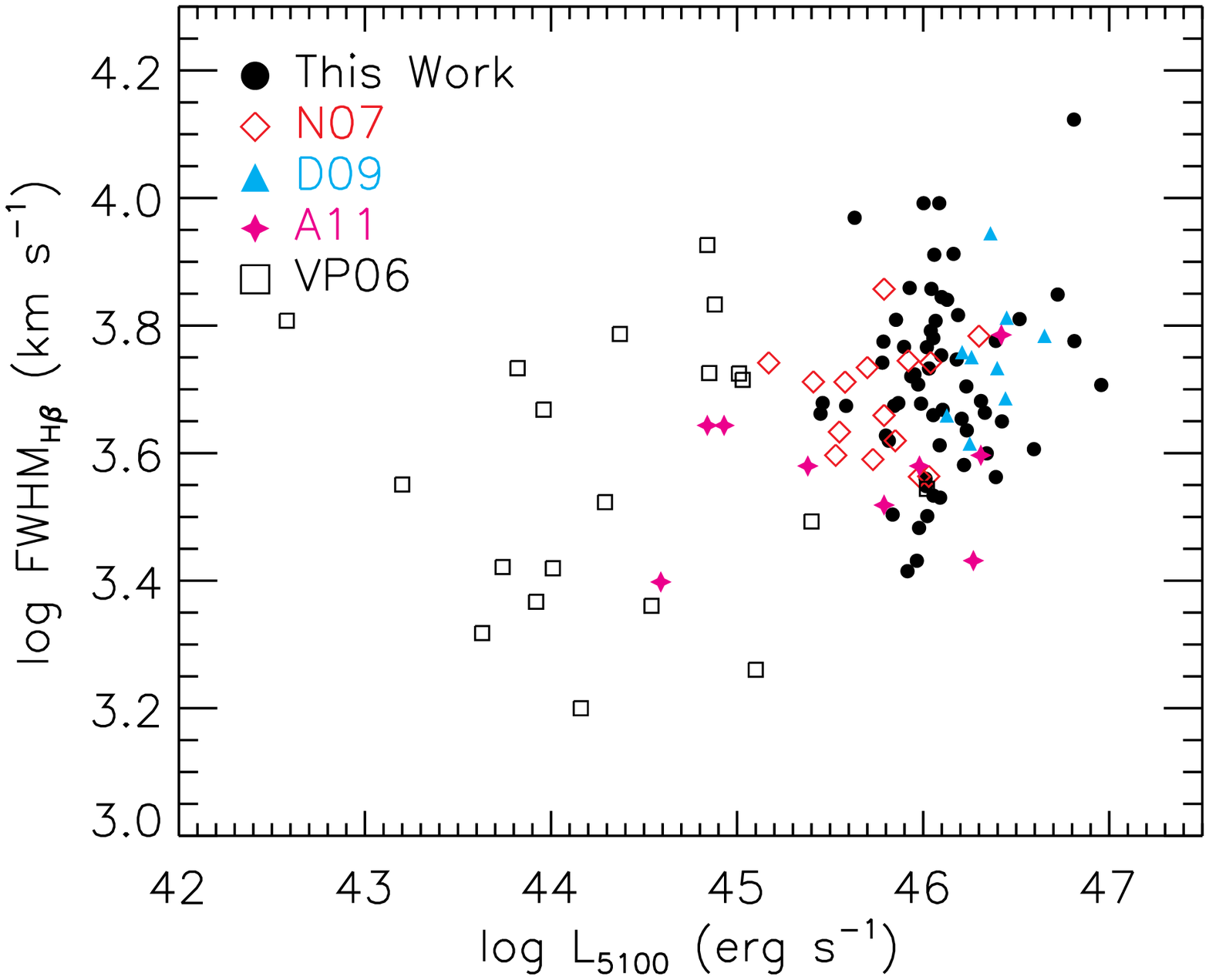}
 \includegraphics[width=0.48\textwidth]{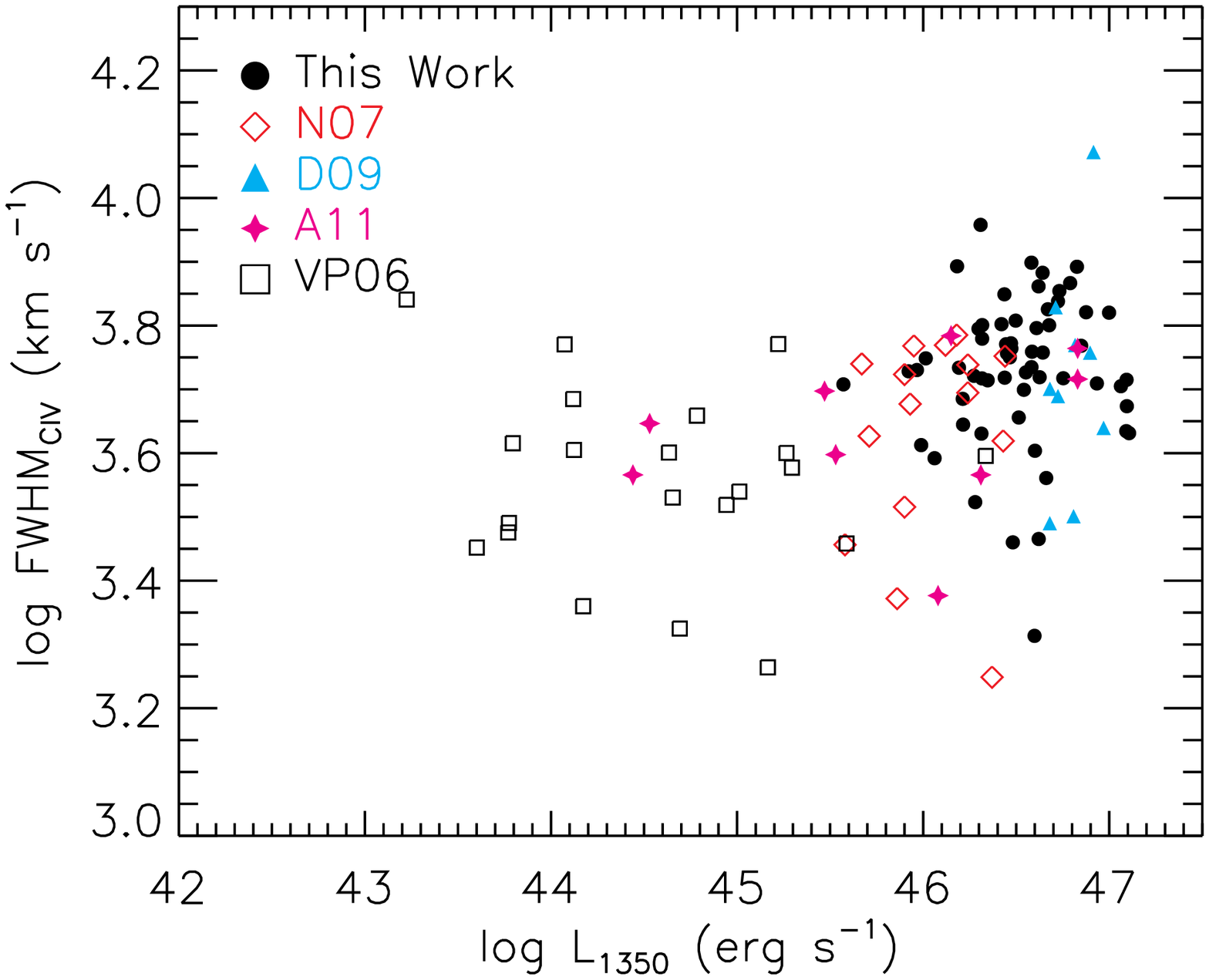}
 \includegraphics[width=0.48\textwidth]{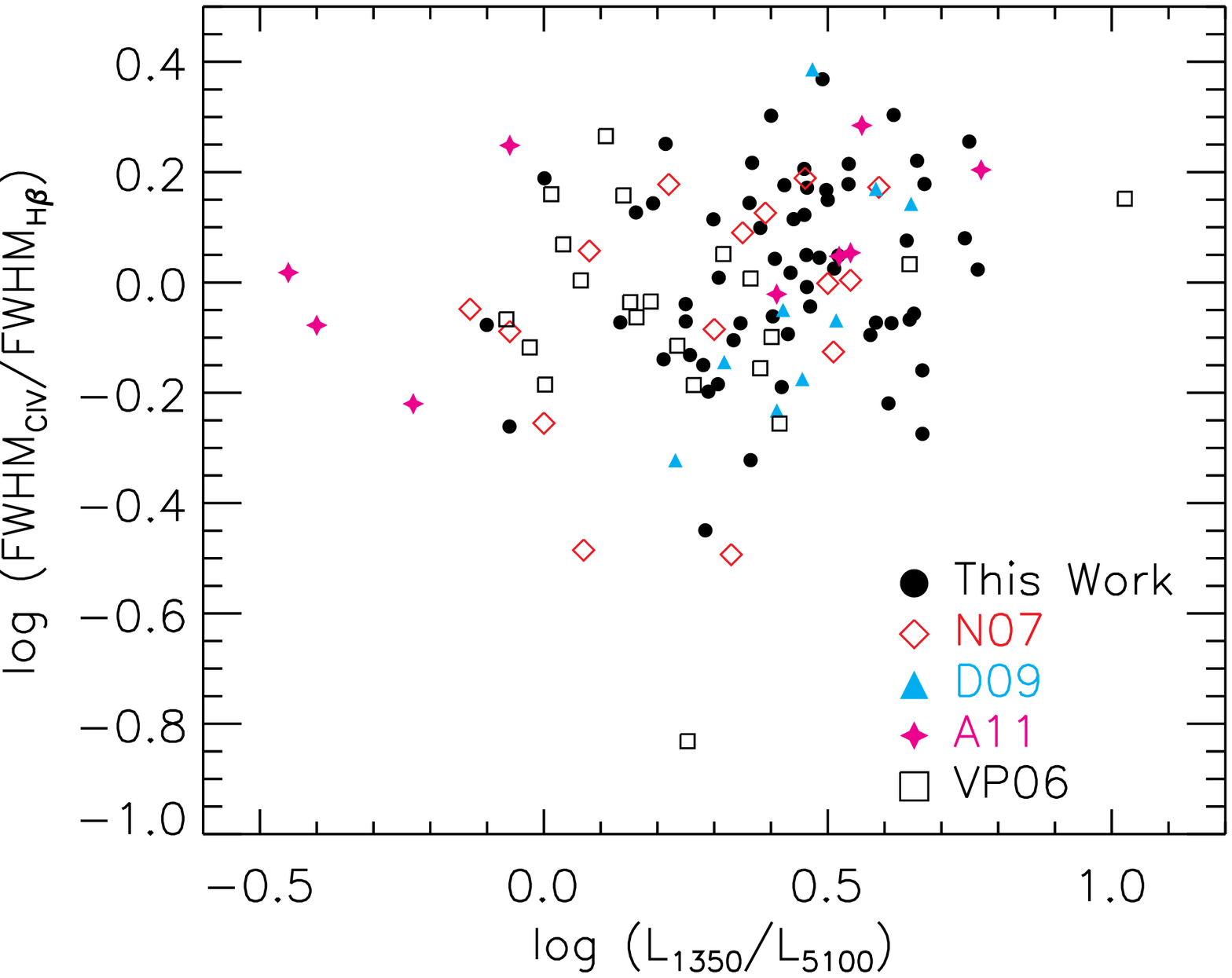}
    \caption{Comparisons between our sample and other samples in the literature [\citet[][9
objects; A11]{Assef_etal_2011}, \citet[][21 objects;
VP06]{Vestergaard_Peterson_2006}, \citet[][15 objects;
N07]{Netzer_etal_2007}, and \citet[][9 objects; D09]{Dietrich_etal_2009}].
    {\em Upper}: Distribution of \hbeta\ FWHM versus $L_{5100}$. {\em Middle:} Distribution
    of \CIV\ FWHM versus $L_{1350}$. {\em Bottom:} Distribution of the ratio
    ${\rm FWHM_{CIV}/FWHM_{H\beta}}$ versus continuum color $L_{1350}/L_{5100}$. Note that
    the VP06 sample has a much fainter luminosity than the other samples. This low-redshift
    sample also has smaller \hbeta\ and \CIV\ FWHMs than the other high-redshift
    samples. Only for the samples in \citet[][9 object]{Dietrich_etal_2009} and \citet[][9 objects]{Assef_etal_2011} is there a significant
    correlation between ${\rm FWHM_{CIV}/FWHM_{H\beta}}$ and $L_{1350}/L_{5100}$, which helps to
    reduce the virial mass difference between \CIV\ and \hbeta\ once this color-dependence is
    taken out.
     }\label{fig:fwhm_comp}
\end{figure}

\begin{figure}
 \centering
 \includegraphics[width=0.48\textwidth]{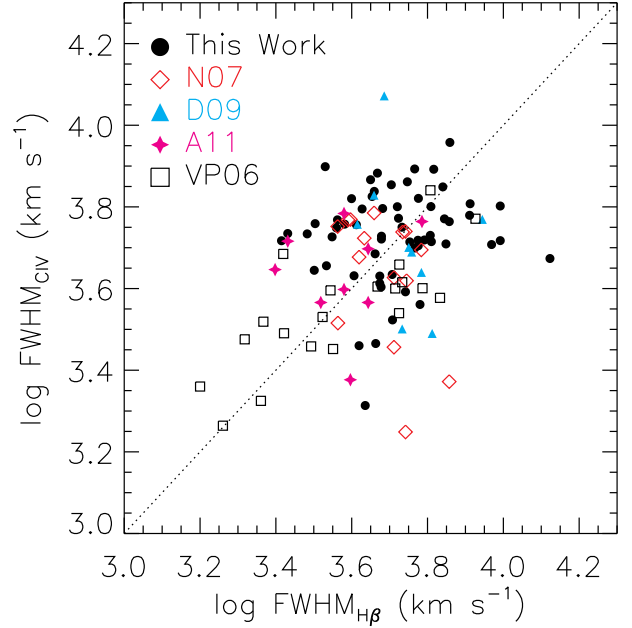}
    \caption{Comparison between \CIV\ FWHM and \hbeta\ FWHM for different samples. The dotted line is the unity relation.
    Only for the low-redshift and low-luminosity VP06 sample is there a significant correlation
    between the two FWHMs. }\label{fig:fwhm_comp_civ_hb}
\end{figure}

The \citet{Assef_etal_2011} sample is small (only 9 objects with all the
measurements available for Spearman tests), and they do not probe a large
dynamic range in \CIV\ blueshift (see their fig.\ 13). The lack of large
\CIV\ blueshift objects in their sample probably explains why they did not
detect a significant correlation between the \CIV\ virial mass residual and
the \CIV\ blueshift.


To further investigate the disagreement on the correlation between \CIV\ and
\hbeta\ FWHMs, we collected luminosity and FWHM measurements from \citet[][9
objects; A11]{Assef_etal_2011}, \citet[][21 objects;
VP06]{Vestergaard_Peterson_2006}, \citet[][15 objects;
N07]{Netzer_etal_2007}, and \citet[][9 objects; D09]{Dietrich_etal_2009}. We
use the Prescription A measurements of \CIV\ FWHM in \citet{Assef_etal_2011}.
Fig.\ \ref{fig:fwhm_comp} shows their distribution in the luminosity-FWHM
space. The VP06 sample probes a much lower luminosity regime than the other
high-redshift samples. In the bottom panel of Fig.\ \ref{fig:fwhm_comp} we
also show the ratio of ${\rm FWHM_{CIV}/FWHM_{H\beta}}$ against the continuum
luminosity ratio $L_{1350}/L_{5100}$ (optical-UV color). The A11 sample
spreads over a larger range in continuum color than the other samples,
including several objects that are much redder than typical quasars
\citep[e.g.,][]{Richards_etal_2003}. Only the D09 and A11 samples show a mild
correlation between FWHM ratio and optical-UV color, with Spearman rank-order
coefficients of 0.65 ($P_{\rm ran}=0.02$) and 0.67 ($P_{\rm ran}=0.05$),
respectively. This correlation helps to reduce the virial mass differences
between \CIV\ and \hbeta\ once this color effect is taken out
\citep{Assef_etal_2011}.

In Fig.\ \ref{fig:fwhm_comp_civ_hb} we show the comparison between \CIV\ and
\hbeta\ FWHMs for different samples. We run Spearman tests for the combined
sample and for each subsample, and find that the correlation between the
\CIV\ and \hbeta\ FWHMs reported in \citet{Assef_etal_2011} is essentially
driven by objects in the VP06 sample, which probes a much fainter luminosity
than other samples as indicated in Fig.\ \ref{fig:fwhm_comp}. None of the
other samples show significant correlations between the two FWHMs, and this
result does not change when we restrict ourselves to high-quality
measurements. These high-redshift samples have a narrower dynamic range in
line width than the VP06 sample, and thus the intrinsic scatter between the
\CIV\ and \hbeta\ FWHMs can easily wash out any weak correlation. We
therefore reinforced our earlier conclusion in \S\ref{subsec:corr_line} that,
at least for the high-luminosity objects, the \CIV\ FWHM is poorly correlated
with the \hbeta\ FWHM.

\subsection{Implications for High-Redshift Quasars}\label{subsec:implic}

The comparisons between different line estimators in previous sections
suggest that in the absence of Balmer lines, the \MgII\ estimator can be used
as a substitute, which will yield consistent virial mass estimates to those
based on the Balmer lines. On the other hand, \CIV\ and \CIII\ can be used,
although using the individually measured \CIV/\CIII\ FWHM does not seem to
offer much advantage over simply using a constant value; of course, this
conclusion is valid for the high-luminosity regime probed by this study.
\CIV\ is a complicated line, and may be more affected by a non-virial
component as luminosity increases \citep[see discussions
in][]{Richards_etal_2011}. These unusual properties of \CIV\ suggest that it
is likely the least reliable virial mass estimator at high-redshift, thus
optical/near-IR coverage of \MgII\ or the Balmer lines is desired for
reliable virial mass estimates
\citep[e.g.,][]{Netzer_etal_2007,Trakhtenbrot_etal_2011,Marziani_Sulentic_2011}.

One should also be aware that even for the most reliable \hbeta-based virial
mass estimates, there is still considerable scatter between virial masses and
true masses \citep[on the level of $> 0.3$\ dex, e.g.][]{Peterson_2011}. Rare
objects with unusual continuum and/or emission line properties \citep[for
instance, dust reddened quasars, e.g., ][]{Richards_etal_2003} will lead to
additional uncertainty in these virial mass estimates, along with measurement
errors from poor spectral quality. Recognizing and accounting for the
uncertainties in these virial mass estimates is crucial in essentially all BH
mass related studies
\citep[e.g.,][]{Shen_etal_2008b,Kelly_etal_2009a,Kelly_etal_2010,Shen_Kelly_2010,Shen_Kelly_2011}.

\section{Summary}\label{sec:con}

In this paper we have empirically determined the relations between
single-epoch virial mass estimators based on different lines for luminous
($L_{5100}>10^{45.4}\,{\rm erg\,s^{-1}}$) quasars, using a sample of 60
intermediate-redshift quasars with complete coverage from \CIV\ through
\halpha\ with good optical and near-IR spectroscopy. Our sample consists of
typical quasars with no peculiarities in their continuum and emission line
properties, has negligible contamination from host starlight, and is large
enough to draw statistically significant conclusions. The main conclusions of
this paper are the following:
\begin{itemize}

\item The \MgII\ FWHM is well correlated with the FWHM of the Balmer
    lines up to high luminosities ($L_{5100}>10^{45.4}\,{\rm
    erg\,s^{-1}}$), which justifies the usage of \MgII\ (in combination
    with $L_{3000}$ or $L_{\rm MgII}$) to estimate virial BH masses for
    luminous quasars at high redshift.

\item The narrow-line contribution to the \CIV\ line is generally
    negligible for high-luminosity quasars; and the FWHMs of \CIV\ and
    \CIII\ are well correlated, suggesting that both lines originate from
    similar regions.

\item The FWHM of \CIV\ is poorly correlated with that of the Balmer
    lines, suggesting different BLRs for the \CIV\ line and for the
    Balmer lines. Part of the discrepancy between the \CIV\ and \hbeta\
    FWHMs is correlated with the blueshift of \CIV\ relative to \hbeta.

\item Using the FWHM of \CIV\ increases the scatter between \CIV\ and
    \hbeta\ based virial masses, which is at least true for the high
    luminosity regime probed in this study. \CIII\ does not seem to be a
    superior substitute for \CIV, both because of the non-correlation
    between \CIII\ and \hbeta\ FWHMs, and because \CIII\ is blended with
    \SiIII\ and \AlIII.

\end{itemize}

While in this work we focused on empirical relations, the correlations (and
lack thereof) among these broad lines and continuum luminosities are
ultimately determined by the accretion disk and BLR physics. In future work,
we will use the same sample to investigate in more detail the emission line
and continuum properties of intermediate-redshift quasars, such as the
Baldwin effect \citep{Baldwin_1977}, line centroid shifts, as well as average
properties and correlations therein, in order to understand the underlying
physics. In the mean time, we will expand our sample to include less luminous
objects and test these correlations over a larger dynamic range in quasar
luminosity.

\acknowledgements We thank Michael Strauss, Benny Trakhtenbrot, Roberto
Assef, Kelly Denney, Chris Kochanek, Brad Peterson, and Jenny Greene for
insightful comments on the manuscript. We also thank Brandon Kelly, Jenny
Greene, and Gordon Richards for useful discussions during the course of the
work, and Rob Simcoe for help with the FIRE data reduction software. Y.S.
acknowledges support from the Smithsonian Astrophysical Observatory through a
Clay Postdoctoral Fellowship. Support for the work of X.L. was provided by
NASA through Einstein Postdoctoral Fellowship grant number PF0-110076 awarded
by the Chandra X-ray Center, which is operated by the Smithsonian
Astrophysical Observatory for NASA under contract NAS8-03060.

Funding for the SDSS and SDSS-II has been provided by the Alfred P. Sloan
Foundation, the Participating Institutions, the National Science Foundation,
the U.S. Department of Energy, the National Aeronautics and Space
Administration, the Japanese Monbukagakusho, the Max Planck Society, and the
Higher Education Funding Council for England. The SDSS Web Site is
http://www.sdss.org/.

Facilities: Sloan, Magellan: Baade (FIRE), ARC 3.5m (TripleSpec)


\end{document}